\documentclass[a4paper,reqno]{amsart}
\usepackage[text={5.0in,8in},centering]{geometry}
\usepackage{amsrefs}
\usepackage{amssymb}
\usepackage{mathrsfs}

\usepackage{xspace}

\usepackage{pgf}
\usepackage{tikz}

\usetikzlibrary{arrows,automata}

\usetikzlibrary{calc}

\DeclareMathAlphabet{\mathpzc}{OT1}{pzc}{m}{it}

\usepackage{epsfig}
\usepackage{enumerate}
\usepackage{graphicx}

\numberwithin{equation}{section}
\theoremstyle{plain}
\newtheorem{theorem}{Theorem}[section]

\newtheorem{lemma}[theorem]{Lemma}
\newtheorem{corollary}[theorem]{Corollary}
\theoremstyle{remark}

\newtheorem*{quest*}{Question}
\newtheorem*{remark*}{Remark}
\theoremstyle{remark}

\theoremstyle{definition}
\newtheorem{definition}{Definition}[section]
\newtheorem*{definition*}{Definition}
\newtheorem*{notation*}{Notation}
\newtheorem*{notations*}{Notations}
\providecommand{\B}{\mathbf}

\providecommand{\D}{\mathbb}

\providecommand{\R}{\mathrm}

\newcommand{\eu}{\mathrm{e}}
\newcommand{\ii}{\mathrm{i}}

\renewcommand{\Im}{\operatorname{Im}}

\def\ball{\mathrm{B}}
\def\bball{\mathbf{B}}

\DeclareMathOperator{\const}{const}
\DeclareMathOperator{\dist}{dist}

\DeclareMathOperator*{\essup}{ess\,sup}
\DeclareMathOperator*{\supp}{supp}

\DeclareMathOperator{\one}{\mathbf{1}}

\DeclareMathOperator{\mes}{{\rm mes}}

\DeclareMathOperator{\diam}{{\rm diam}}

\def\rc{\mathrm{c}}

\def\lam{{\lambda}}

\def\eps{\epsilon}

\def\PI{{\rm PI}\xspace}
\def\FI{{\rm FI}\xspace}

\def\SSI#1{\textsf{S}$(I,#1)$}

\def\ER{$E$-R\xspace}

\def\EmS{$(E,m)$-S\xspace}
\def\EmNS{$(E,m)$-NS\xspace}

\def\ET{$E$-T\xspace}
\def\ENT{$E$-NT\xspace}
\def\ER{$E$-R\xspace}
\def\ENR{$E$-NR\xspace}

\def\tE{{ \widetilde{E} }}
\def\tK{{ \widetilde{K} }}

\def\htil{{\widetilde{h}}}

\def\brho{{\boldsymbol{\rho}}}
\def\brhoS{{\boldsymbol{\rho}_{\R{S}}}}

\def\hBx{\widehat{\Bx}}

\def\BP{\B{P}}

\providecommand{\bS}[1]{\boldsymbol{#1}}

\def\BA{\mathbf{A}}
\def\BF{\mathbf{F}}
\def\BG{\mathbf{G}}
\def\BH{\mathbf{H}}
\def\BHN{\mathbf{H}^{(N)}}

\def\BU{\mathbf{U}}
\def\BV{\mathbf{V}}
\def\BbS{\mathbf{S}}
\def\Bx{\mathbf{x}}
\def\By{\mathbf{y}}
\def\Bu{\mathbf{u}}

\def\BDelta{{\boldsymbol{\Delta}}}
\def\Bpsi{{\boldsymbol{\psi}}}
\def\Bphi{{\boldsymbol{\phi}}}
\def\PPi{{\mathrm{\Pi}}}

\def\BLam{\mathbf{\Lambda}}

\def\BPsi{{\bS{\Psi}}}

\def\Const{{\rm{Const}}}

\def\tn{{\widetilde{n}}}

\def\DC{\D{C}}

\def\DP{\D{P}}
\def\DR{\D{R}}
\def\DZ{\D{Z}}
\def\DN{\D{N}}

\def\cB{\mathcal{B}}
\def\csA{\mathscr{A}}
\def\csB{\mathscr{B}}
\def\csE{\mathscr{E}}
\def\csO{\mathscr{O}}
\def\csR{\mathscr{R}}

\def\cD{\mathcal{D}}
\def\cE{\mathcal{E}}

\def\cG{\mathcal{G}}

\def\cJ{\mathcal{J}}

\def\cM{\mathcal{M}}
\def\cN{\mathcal{N}}
\def\cP{\mathcal{P}}
\def\csP{\mathscr{P}}
\def\csQ{\mathscr{Q}}

\def\cS{\mathcal{S}}

\def\cX{\mathcal{X}}

\def\cZ{\mathcal{Z}}

\def\cN{{\mathcal{N}}}

\def\boxx{{\mathrm B}}

\def\brd{{\boldsymbol{\R{d}}}}
\def\rD{{\R{D}}}
\def\rN{{\R{N}}}
\def\rd{{\R{d}}}
\def\rP{{\R{P}}}
\def\rQ{{\R{Q}}}
\def\rS{{\R{S}}}

\def\hN{\hat{N}}

\def\be{\begin{equation}}
\def\ee{\end{equation}}
\def\ba{\begin{array}{l}}
\def\ea{\end{array}}
\def\bal{\begin{aligned}}
\def\eal{\end{aligned}}

\def\fF{\mathfrak{F}}
\def\fS{\mathfrak{S}}

\def\om{{\omega}}
\def\Om{{\Omega}}

\def\eps{\epsilon}

\def\Lam{{\Lambda}}
\def\lam{{\lambda}}

\def\bcG{{\boldsymbol{\mathcal{G}}}}

\def\bcZ{{\boldsymbol{\mathcal{Z}}}}
\def\bcZN{{\boldsymbol{\mathcal{Z}^{N}}}}

\def\bcEN{{\boldsymbol{\mathcal{E}^{(N)}}}}

\def\pr#1{\D{P}\left\{\,#1\,\right\}}
\def\esm#1{\D{E}\left[\, #1\, \right]}
\def\pt{\partial}
\def\half{\frac{1}{2}}

\def\nb#1{{ \langle #1 \rangle}}

\def\tto#1{\smash{\mathop{\,\,\,\, \longrightarrow \,\,\,\, }\limits_{#1}}}

\def\myset#1{{\left\{\,#1\,\right\}}}

\def\Wone{{\bf (W1)}}
\def\Wtwo{{\bf (W2)}}

\def\Uone{\textsf{(U1)}\xspace}

\def\Wone{\textsf{(W1)}\xspace}
\def\Wtwo{\textsf{(W2)}\xspace}

\def\DeltaD{\Delta^{\rm D}}

\begin{document}

\title[Fixed-energy multi-particle MSA implies dynamical localization]
{Fixed-energy multi-particle MSA\\ implies dynamical localization}

\author[V. Chulaevsky]{Victor Chulaevsky}


\address{D\'{e}partement de Math\'{e}matiques\\
Universit\'{e} de Reims, Moulin de la Housse, B.P. 1039\\
51687 Reims Cedex 2, France\\
E-mail: victor.tchoulaevski@univ-reims.fr}

\date{}
\begin{abstract}
This work is a continuation of \cite{C12b} where we described two elementary
derivations of the variable-energy MSA bounds from their fixed-energy counterparts,
in the framework of single-particle disordered quantum particle systems on graphs
with polynomially bounded growth of balls.
Here the approach of \cite{C12b} is extended to
multi-particle Anderson Hamiltonians with interaction; it plays a role similar
to that of the Simon--Wolf criterion for single-particle Hamiltonians.
A simplified,
fixed-energy multi-particle MSA scheme was developed
in our earlier work \cite{C08a}, based on a multi-particle adaptation of
techniques from Spencer's paper \cite{Sp88}. Combined with a simplified variant of the Germinet--Klein argument \cite{GK01} described in \cite{C12a}, the outcome of the
fixed-energy analysis results in an elementary proof of multi-particle
dynamical localization with the decay of eigenfunction correlators
faster than any power-law.
\end{abstract}

\maketitle

\section{Introduction. } \label{sec:intro}

\vskip2mm

This manuscript is an extended version of my talk given recently at the workshop
"\emph{Mathematics of quantum disordered systems}" organized by
the \emph{Institut de Math\'{e}matiques de Jussieu} at Chevaleret
(Paris) and by the \emph{Institut Galil\'{e}e} at Villetaneuse.
A reader familiar with \cite{C12b} may notice that substantial portions
of the text are borrowed from that preprint, often by the "copy-n-paste"
method. A good time-saving tool, it is also a well-known source of a.s. creation
of notational inconsistencies, for which I sincerely apologize in advance.
In my talk, this text has been promised to be uploaded promptly
to \texttt{arXiv}; a more detailed version will be made available later.

I would like to stress
here a point which may have been insufficiently emphasized in the talk: although 
several
approaches to the derivation of the spectral and dynamical localization from the
FEMSA (fixed-energy multi-scale analysis) estimates for \textbf{single-particle}
systems are known by
now\footnote{I thank Peter Hislop for a fruitful exchange on this subject
during the above mentioned workshop "\emph{Mathematics of quantum disordered systems}".}
(cf., e.g.,
\cite{BK05}, \cite{GK11}), this text focuses  only on two methods which seem
to provide the shortest path to the multi-particle localization from the FEMPMSA
results,
which are \underline{\textbf{much}} simpler to obtain than their variable-energy analogs,
the VEMPMSA (variable-energy multi-particle MSA) bounds.
The choice is certainly biased by the author's personal preferences, and there is no doubt
that alternative, more efficient and more general methods will appear in near future.

Several ideas employed in the derivation
"FEMPMSA $\Rightarrow$ VEMPMSA" have been used by other researchers in
the context of single-particle systems.  The
"disorder-energy" measurable space $\Om\times\DR$ appears already in the work by Martinelli--Scoppola \cite{MS85}. In a more general context, it
becomes the main scene of action in the Simon--Wolf paper \cite{SW86}
(see \cite{SW86} for a more extensive bibliography and a
discussion of a series of works which initiated the research
by Simon and Wolf).
%
The fact that the matrix elements of resolvents are rational functions with a "moderate" number of poles is explicitly used, e.g., by Bourgain--Kenig \cite{BK05}, in a
"hard" situation (Bernoulli--Anderson Hamiltonians). Germinet--Klein \cite{GK11}
use spectral
reductions for random Hamiltonians with virtually no assumption on the marginal
probability distribution of the random potential (the scatterers amplitude of an
alloy potential may follow any probability law not concentrated on a single point).
While the singular nature of the random potential forces one to make use of
technically involved analytic and probabilistic tools, making a strong assumption
on the regularity of the marginal distributions (viz., sufficient regularity
of the probability \emph{density}) gives rise to a significant simplification
of spectral reductions (FEMSA $\Rightarrow$ VEMSA).

This list  can be continued ...

Needless to say that the Fractional Moment Analysis (FMM)
always starts as a fixed-energy analysis.

In the framework of the multi-particle Anderson Hamiltonians, the above mentioned strong
assumption on the regularity of the marginal \emph{density} remains so far
the only means to achieve more optimal, physically reasonable bounds on long-range
charge transfer processes (tunneling) in an interacting quantum system of
$N\ge 3$ particles; cf. a brief discussion in
subsection \ref{ssec:assump.V}. Resonances occurring in
such systems can be qualified as "structural"; they do not appear in $1$-particle
systems and can be treated in a relatively simple way in $2$-particle
systems. The solution to this problem (explicitly
analyzed by Aizenman and Warzel \cite{AW09a}), proposed in our works \cite{C10}, \cite{C11a},
requires the above mentioned regularity assumption. Therefore, limiting the spectral
reduction (FEMSA $\Rightarrow$ VEMSA) to random Hamiltonians obeying this
assumption does not seem to be an overly big concession -- at least, until a new, more
efficient solution is found to the problem of long-range tunneling due to
structural resonances in systems with $N>2$ particles.

On the other hand, proving merely the Anderson localization phenomenon in an interacting
quantum system with an arbitrary (but fixed) finite number of particles is a simpler
task than proving \emph{efficient} decay bounds for eigenfunction
correlators. This can be done under a much weaker hypothesis of
H\"{o}lder (or even log-H\"{o}lder) continuity of the \emph{cumulative} marginal
probability distribution function (PDF) of the external random potential featuring
the IID, IAD (=Independence At Distance) or an (appropriate) strong mixing
property.
The main geometrical
tool here is the notion of "separability" of pairs of finite volumes
developed in our joint works with Yuri Suhov \cite{CS08,CS09a} (for $N=2$
particles) and in \cite{CS09b} (for $N\ge 2$ particles).

\section{Basic notations, facts and assumptions}
\label{sec:notations}

Throughout this paper, we work with discrete Schr\"{o}dinger operators (DSO) acting in
Hilbert spaces of square-summable complex functions on connected countable graphs.
Indeed, the techniques and results of the MSA, initially developed for operators on periodic lattices, are naturally extended to more general graphs with polynomially bounded growth of
balls (such graphs as Bethe lattices remain so far out of the MSA's reach). Another motivation for
presenting the new approach on a graph comes from the fact that the natural language
for the description of a system of $N>1$ interacting indistinguishable quantum particles
(bosons or fermions) is that of a symmetric power of the configuration space  $\cZ$ of the
respective single-particle system; already in the case where the configuration space
is $\cZ=\DZ^d$, $d>1$, its $N$-th \emph{symmetric} power is no longer a periodic lattice.

\subsection{Graphs, configurations and graph Laplacians}

Consider a finite or countable connected graph $(\cG, \cE)$, with the set of vertices $\cG$ and the set of edges $\cE$; for brevity, we will often call $\cG$ the graph, omitting the reference to $\cE$. We denote by $\rd_\cG(\cdot,\,\cdot)$ (sometimes simply by $\rd(\cdot\,,\cdot)$)
the canonical distance on the graph $\cG$: $\rd_\cG(x,y)$ is the length of the shortest path
$x\rightsquigarrow y$ over the edges.
We will assume that the growth of balls $\ball_L(x):=\{y:\, \rd_\cG(x,y)\le L\}$ is polynomially bounded:
\be\label{eq:ball.growth}
\sup_{x\in\cG}| \ball_L(x)| \le C_d L^d, \;\; L\ge 1.
\ee
In particular, the coordination number $n_\cG(x) :=\{y:\, \rd_\cG(x,y)=1\}$ of any vertex $x$
is bounded by $C_d$ (even by $C_d - 1$).

Given a connected graph $(\cZ,\cE_\cZ)$ serving as the configuration space of quantum particles,
the configuration space of a system of $N>1$ distinguishable particles is the
cartesian product $\cZ^N$; it is usually endowed with the following graph structure:
a pair $(\Bx, \By)$ is an edge iff, for some $j_\circ\in[1,N]$,
$(x_{j_\circ},y_{j_\circ})\in\cE_\cZ$, while for all $i\ne j_\circ$ one has $x_i = y_i$.
In other words, $\By$ is obtained from $\Bx$ by moving exactly one particle
to one of its nearest neighbors in $\cZ$. To make explicit this choice, we use the
boldface notations $\bcZN$ for the vertex set of the $N$-particle
configuration graph and $\bcEN$ for the described edge set. In general, boldface notations
will be reserved for "multi-particle" objects.

Introduce the following mapping from $\cZ^n$, $n\ge 1$, to the collection
of finite subsets of $\cZ$:
\be\label{eq:def.Pi}
\Pi: \Bx = (x_1, \ldots, x_n) \mapsto \{x_1, \ldots, x_n\}.
\ee
We will call $\Pi\Bx$ \emph{the support} of the configuration $\Bx\in\cZ^n$.
(Note that in the framework of indistinguishable particles, only $\Pi \Bx$
would be physically observable.) Similarly, the support
of a "polydisk" $\bball^{(n)}_L(\Bx) = \times_{j=1}^n \ball_L(x_j)$ is the set
\be\label{eq:def.Pi.ball}
\Pi \bball_L(\Bx) = \cup_{j=1}^n \ball_L(x_j) \subset\cZ.
\ee
Further, given a non-empty index subset $\cJ\subseteq[1,N]\cap \DZ$, define
a partial projection (or partial support)
$$
\Pi_\cJ: (x_1, \ldots, x_N) \mapsto \{x_j, j\in\cJ\}\subset \cZ.
$$
For $\cJ=\varnothing$ set, formally, $\Pi_\varnothing \Bx = \varnothing$.

Apart from the graph distance $\brd(\cdot\,,\,\cdot) = \rd_{\bcZN}(\cdot\,,\,\cdot)$
on $\bcZN$, it will be convenient to use the max-distance $\brho$
and symmetrized max-distance $\brhoS$ defined as follows:
$$
\brho(\Bx, \By) = \max_{1\le j \le N} \rd_{\cZ}(x_j, y_j), \quad
\brhoS(\Bx, \By) = \min_{\pi\in\fS_N} \brho\big(\Bx, \pi(\By) \big),
$$
where the elements $\pi\in\fS_N$ of the symmetric group $\fS_N$
act on vertices $\Bx\in\cZ^N$ by permutations
of the coordinates.
In terms of $\brho$, a polydisk $\bball^{(n)}_L(\Bx)$ is a ball of radius $L$
centered at $\Bx$. (If $\cZ = \DZ^d$ with max-distance, then polydisks are cubes.)

The canonical (negative) graph Laplacian $(-\Delta_\cG)$ on a finite or countable graph
$(\cG, \cE)$ is given by
\be\label{eq:def.Laplace.graph}
(-\Delta_\cG f)(x) = \sum_{ \nb{x,y} } (f(x) - f(y))
= n_\cG(x) f(x) - \sum_{ \nb{x,y} } f(y)
\ee
where we use a popular notation $\nb{x,y}$ for a pair of nearest neighbors $x,y\in\cG$, i.e., $\rd_\cG(x,y)=1$, and $n_\cG(x)$
is the coordination number of the point $x$. For brevity, we will sometimes use
slightly abusive notations like $\nb{x,y}\in\Lam$, $\Lam\subset\cG$
instead of $\nb{x,y}\in(\Lam\times\Lam) \cap \cE_\cG$.

From this point on, unless otherwise specified, we will use the notation $\cG$ only for
finite connected graphs, either in the single-particle context or in a situation where the
nature of a graph (single- or multi-particle) is irrelevant, while
$\cZ$ will stand for a countable connected graph
with polynomial growth of balls. Finite connected subgraphs
of $\bcZN$ will be sometimes denoted by $\bcG$.

In operator form, we can write, for an arbitrary (connected) graph $\cG$,
$$
-\Delta_\cG = n_\cG - \sum_{ \nb{x,y} } \Gamma_{x,y},
\quad
\Gamma_{x,y} = | \one_x \rangle \langle \one_y |,
$$
where $n_\cG$ is the operator of multiplication by the function $x\mapsto n_\cG(x)$.
Given a subgraph $\Lam\subsetneq \cG$, define its internal, external and
the so-called edge boundary (relative to $\cG$) as follows:
$$
\bal
\pt^-_{\cG} \Lam &= \{y\in\Lam:\, \rd_\cG(x, \cG\setminus \Lam) = 1\},
\quad
\pt^+_{\cG} \Lam = \pt^-_{\cG } \cG\setminus \Lam,
\\
\pt_{\cG} \Lam &= \{(x,y)\in\pt^-_\cG \Lam\times\pt^+_\cG  \Lam: \,\rd_\cG(x, y)=1\}.
\eal
$$

Working with a given graph $\cG (\subset\cZ)$, we always mean by a ball $\ball_R(u)\subset\cG$
the set $\{y\in\cG:\, \rd_\cG(u,y)\le R\}$, i.e., the \textbf{ball relative to the metric space}
$(\cG,\rd_\cG)$.

The Laplacian (hence, a DSO) in a subgraph $\Lam\subset\cG$ can be defined in various ways. The two most popular choices are:
\begin{itemize}
  \item The canonical (negative) Laplacian in $\Lam$, $(-\Delta^\rN_\Lam f)=(-\Delta_\Lam f)$,
  defined as in \eqref{eq:def.Laplace.graph} with $\cG$ replaced by
  $\Lam$. It this context, it is usually considered as an analog of the Neumann Laplacian, and reads as follows:
\be\label{eq:Laplace.N}
(-\Delta^\rN_\Lam f)(x) = n_{\Lam}(x) - \sum_{\nb{x,y}\in \Lam} f(y).
\ee
  \item The Dirichlet Laplacian
$(-\Delta^\rD_{\Lam,\cG}) = \one_{\Lam} (-\Delta^\rD_{\Lam,\cG}) \one_{\Lam} \upharpoonright
\ell^2(\Lam)$. Here we use a natural injection $\ell^2(\Lam)\hookrightarrow \ell^2(\cG)$.
The Dirichlet counterpart of \eqref{eq:Laplace.N} is
\be\label{eq:Laplace.N}
(-\Delta^\rD_\Lam f)(x) = n_{\cG}(x) - \sum_{\nb{x,y}\in \Lam} f(y),
\ee
with $n_\cG(x)\ge n_\Lam(x)$, so
$(-\Delta^\rD_\Lam) \ge (-\Delta^\rN_\Lam)$ in the sense of quadratic forms.
\end{itemize}

We will use the Dirichlet Laplacians and DSO $H^\rD_\Lam$.
Given a decomposition $\cG = \Lam \sqcup \Lam^\rc$, $\Lam^\rc := \cG\setminus\Lam^\rc$,
we can write
$$
\bal
-\DeltaD_\cG &= n_\cG -\sum_{ \nb{x,y}\in\Lam } \Gamma_{x,y} - \sum_{ \nb{x,y}\in\Lam^\rc } \Gamma_{x,y}
- \sum_{ \nb{x,y}\in\pt \Lam } \left(\Gamma_{x,y} + \Gamma_{y,x} \right)
\\
& =  \left( (-\DeltaD_\Lam) \oplus (-\DeltaD_{\Lam^\rc}) \right) - \Gamma_{\Lam,\cG}
\eal
$$
with $\Gamma_{\Lam,\cG}=\sum_{ \nb{x,y}\in\pt \Lam } \left(\Gamma_{x,y} + \Gamma_{y,x} \right)$.
Respectively for the DSO $H_\cG = -\DeltaD_\cG + V$, where
$V:\cG \to \DR$ is usually referred to as the potential, one has
$$
H_\cG
= H^\bullet_{\cG,\Lam} - \Gamma_{\Lam,\cG},
\qquad
H^\bullet_{\cG,\Lam} := (-\DeltaD_\Lam + V) \oplus (-\DeltaD_{\Lam^\rc} + V).
$$
We omit the superscript "$\rN$", since the nature of the boundary conditions in $\cG$
is not related to the choice of Dirichlet or Neumann decoupling
induced by $\cG = \Lam \sqcup \Lam^\rc$.

The spectrum of a (finite-dimensional) operator $H_\cG$, i.e., the set of its eigenvalues (EVs)
counting multiplicities, will be denoted by $\Sigma(H_\cG)$. The resolvent
of a Hamiltonian $\BH_{\BLam}(\om)$, $\BLam\subseteq \bcZN$, will be denoted
by $\BG_\BLam(E; \om)$ and its matrix elements in the delta-basis (Green functions)
by $\BG_\BLam(\Bx, \By; E; \om)$. The subscript $\BLam$ will be omitted when
$\BLam=\bcZN$.

In a number of formulae and statements, we will use parameters $\beta,\tau,\varrho\in(0,1)$,
and $\alpha\in(1,2)$. Unless otherwise specified, we assume that $\beta=1/2$, $\tau=1/8$,
$\varrho = (\alpha-1)/2 = 1/6$ and $\alpha = 3/2$. Note that the exponent
$\frac{1+\varrho}{\alpha}$ figuring in Definition \ref{def:NR.NS} then equals $7/8$.

The function $\gamma: (m,L)\mapsto m(1+L^{-\tau})$ introduced below
is a convenient replacement for the decay exponent ("mass") used in the MSA and dependent
upon the scale. We merely make this dependence explicit. Clearly, $\gamma(m,L)>m$
for any $L>0$.

\begin{definition}\label{def:NR.NS}
Given numbers $E\in\DR$, $m>0$ and $L\in\DN^*$, a ball $\bball_L(\Bu)$ is called
\begin{itemize}
  \item $E$-resonant (\ER, in short), if
  $\dist(\Sigma(\BH_{\bball_L(\Bu)}), E) < \eu^{-L^\beta}$,
  and $E$-nonresonant (\ENR), otherwise;
  \item $(E,m)$-nonsingular (\EmNS), if for all $\Bx,\By\in\bball_L(\Bu)$ with
  $\brho(\Bx,\By)\ge L^{\frac{1+\varrho}{\alpha}}$
\be\label{eq:def.NS}
| \pt\bball_L(\Bu)| \cdot
|\BG_{\bball_L(\Bu)}(\Bx,\By;E)| \le \eu^{-\gamma(m,L)\brho(\Bx,\By)},
\ee
where
\be
\gamma(m,L) := m(1+L^{-\tau}),
\ee
and $(E,m)$-nonsingular (\EmNS), otherwise.
\end{itemize}

\end{definition}


\subsection{Geometric resolvent inequality}

The second resolvent identity implies the so-called Geometric resolvent equation for the resolvents $G_\cG(E) = (H_\cG - E)^{-1}$, $G_{\Lam^\rc}(E) = (H_{\Lam^\rc} - E)^{-1}$,
$G^\bullet_{\cG,\Lam}(E) = (H^\bullet_{\cG,\Lam} - E)^{-1}$:
\be\label{eq:GRE.resolvents}
G_\cG(E) = G^\bullet_{\cG}(E) + G^\bullet_{\cG}(E) \, \Gamma_{\Lam,\cG} \, G_\cG(E).
\ee
For $x, u\in\Lam$ and $y\in\Lam^\rc$, one has
$G^\bullet_{\cG}(x,u;E)=G^\rD_{\Lam}(x,u;E)$ and
$G^\bullet_{\cG}(x,y;E)=0$. This results in the Geometric resolvent equation
for the Green functions
\be\label{eq:GRE}
G_\cG(x,y;E) = \sum_{ \nb{u,u'}\in\pt_\cG \Lam} G^\rD_{\Lam}(x,u;E)\, G_{\cG}(u',y;E)
\ee
and the Geometric resolvent inequality (GRI)
\be\label{eq:GRI}
|G_\cG(x,y;E)| \le
\sum_{ \nb{u,u'}\in\pt_\cG \Lam} |G^\rD_{\Lam}(x,u;E)|\, |G_{\cG}(u',y;E)|.
\ee

In the rest of the paper, the superscript "D" will be omitted, since we do not use the Neumann
boundary conditions.

\subsection{Assumptions on the random potential.}
\label{ssec:assump.V}

The efficiency of the quantitative bounds on eigenfunctions and eigenfunction correlators
for $N\ge 3$ interacting particles available at the moment
and the complexity of the proofs depend upon the assumptions
on the random potential.

For clarity of presentation, we always assume that the random potential field
$V:\cZ\times\Om\to\DR$ on a graph $\cZ$ is IID.

The first, more general condition (leading to weaker results) is as follows:
\par\medskip
\Wone:
\textit{ The marginal probability distribution function (PDF) $F_V$ is uniformly
H\"{o}lder continuous: there are constants $C_H\in(0,+\infty)$, $\delta\in(0,1]$
such that
\be\label{eq:Holder}
\sup_{t\in\DR} (F_V(t+s) - F_V(t)) \le C_H s^{\delta}.
\ee
}

To obtain more optimal decay bounds on eigenfunction correlators,
with the help of a method developed in \cite{C11a}--\cite{C10}, one needs
an additional assumption which we will describe now.
Introduce the following notations. Given a finite subset  $\Lam\subset\cZ$, let
$\xi_{\Lam}(\omega)$ be the sample mean of the random field $V$ over the $\Lam$,
$$
 \xi_{\Lam}(\omega)  = | \Lam |^{-1} \sum_{x\in \Lam} V(x,\omega)
$$
and define the "fluctuations" of $V$ relative to the sample mean,
$
 \eta_x(\om)  = V(x,\omega) - \xi_{\Lam}(\omega), \; x\in \Lam.
$
Denote by $\fF_{\eta,\Lam}$ the sigma-algebra generated by
$\{\eta_x:\, \,x\in \Lam\}$, and by
$F_{\xi_\Lam}( \cdot\,| \fF_{\eta,\Lam})$ the conditional distribution function of
$\xi_\Lam$ given $\fF_{\eta,\Lam}$.
Assuming that $\Lam\subset\cZ$ and $\diam(\Lam)\le R$,
introduce, for $s\ge 0$, an
$\fF_{\eta,\Lam}$-measurable random variable
\begin{equation}\label{eq:def.nu.R}
\nu_{\xi_\Lam}(s;\om) :=
\essup\; \sup_{t\in\DR}    |F_{\xi_\Lam}(t+s\,| \fF_{\eta,\Lam}) - F_{\xi_\Lam}(t\,| \fF_{\eta,\Lam})|.
\end{equation}

The stronger assumption on the random field $V$ is as follows:
\par\medskip
\Wtwo:
\textit{ There exist $C',C'', A', A'', B', B''\in(0,+\infty)$ such that
for any finite subset $\Lam\subset\cZ$ with $\diam(\Lam)\le R$,
\begin{equation}\label{eq:CMnu}
\forall \, s\in(0,1]\qquad
\pr{ \nu_{\xi_\Lam}(s;\om) \ge C R^{A} s^{B}} \le C' R^{A'} s^{B'}.
\end{equation}
} 

In the case where the random field $V$ is IID, a more
natural condition should refer to the cardinality $|\Lam|$ of the subset
$\Lam$, but we will use the above property only in a situation where
$\Lam$ is a subset of some ball, of an explicitly bounded diameter.

For further use, define the following function on $[0,1]$:
\be\label{eq:h.L.weak.sep}
h_L = h_L^{(N,\Bx,\By)}: s \mapsto
  |\bball^{(N)}_{L}(\Bx)| \cdot |\bball^{(N)}_{L}(\By)| C L^A s^B + C' L^{A'} s^{B'}.
\ee

See the discussion of the validity of \Wtwo in subsection \ref{ssec:validity.W2}
below.

\subsection{Assumptions on the interaction potential.}

For brevity, we consider only finite-range interactions $\BU$ generated by a pair interaction
potential $U^{(2)}(\cdot)$:

\Uone: \textit{ There is an integer $r_0\ge 0$ and a function
$U^{(2)}: \DN \to \DR$ with finite support, $\supp U^{(2)} \subset [0,r_0]$, such that
\be\label{eq:def.U}
\BU(\Bx) = \sum_{1 \le i < j \le N} U^{(2)}\big( \rd(x_i,x_j) \big).
\ee
}

The parameter $r_0$ will be called the range of the interaction $\BU$.

\vskip2mm

In the case where the single-particle configuration space is a group (e.g., $\DZ^d$
or $\DR^d$), the above form of $\BU$ corresponds to translation invariant
interactions; in the case of a Euclidean space, such an interaction is also
isotropic. In all cases, $\BU$ of the form \eqref{eq:def.U}
is permutation symmetric. We stress that neither
of these properties is crucial for the multi-particle MSA scheme. Moreover,
this scheme can be easily adapted to interactions with a hard core.

The assumption of finite range simplifies the induction on particles. In \cite{C11b}, we
described an extension of the variable-energy MPMSA
to interactions decaying at an exponential or subexponential rate,
$$
| U^{(2)} (r)| \le c_1 \eu^{-c_2 r^{c_3}}
$$
with $c_1, c_2\in(0,+\infty)$ and $c_3>0$ sufficiently close to $1$. The proof
 of the fixed-energy MPMSA bounds can follow essentially the same path. In fact, the most significant modification is required in the proof of Lemma \ref{lem:PITRONS}; see the details
in \cite{C11b}. In the framework of the fixed-energy MPMSA (which is simpler),
we plan to address infinite-range interactions in a forthcoming paper.

\section{Main results}
\label{sec:main.results}

Let $\csB_1$ be the set of all bounded Borel functions $f$ with $\|f\|_\infty\le 1$.

\begin{theorem}\label{thm:Main.DL.Wone}
Fix any integer ${\hN}\ge 2$.
Under the assumptions {\rm \Wone} and {\rm\Uone}, for any  $N\in[1,{\hN}]$ there exists $g_0(N)<+\infty$ such that
for all $|g|\ge g_0(N)$ and some $m=m(g,N)\ge Const(N) \ln |g|>0$:
\par\noindent
{\rm{\textbf{(A)}}}
with probability one, the random operator $\BH(\om)=\BH_0 + g\BV(\om)+\BU$
has pure point spectrum and all
eigenfunctions $\BPsi_j(\om)$ rapidly decaying at infinity:  for each $\BPsi_j$
and some $\widehat{\Bx}_j$, for all $\Bx$ and some $C(\hBx_j,\om)\in(0,+\infty)$,
\be\label{eq:thm.Main.DL.1}
| \BPsi_j(\Bx,\om) | \le C(\hBx_j,\om) \,\eu^{ -a\ln^{1+c} \brho(\Bx,\hBx_j) };
\ee
\par\noindent
{\rm{\textbf{(B)}}}
for all points $\Bx,\By$ and some $a,c, C(\Bx)\in(0,+\infty)$,
\be\label{eq:thm.Main.DL.2}
\esm{ \sup_{f\in\csB_1} \big| \langle \one_{\By}\, | \,  f(\BH(\omega)) \, | \, \one_{\Bx}\rangle\big| }
\le C(\Bx) \, \eu^{ -a\ln^{1+c} \brho(\Bx,\By) }.
\ee
\end{theorem}

With $f = f_\lam: t\mapsto \eu^{-it\lam}$, $\lam\in\DR$,
Eqn \eqref{eq:thm.Main.DL.2}
reads as the strong dynamical localization for the random Hamiltonians
$\BH^{(N)}(\om)$.

Technically, we prove upper bounds in \eqref{eq:thm.Main.DL.1}-\eqref{eq:thm.Main.DL.2}
of the form
$L_k^{-\kappa(1+\theta)^k}$ at distances $\sim L_k \sim (L_0)^{\alpha^k}$, which can be easily
translated into the above decay rate.

Assertion (A) can be made stronger, without replacing the hypothesis
\Wone by a stronger condition \Wtwo. Indeed, one can prove an exponential decay
of all eigenfunctions, using the variable-energy MPMSA developed in
\cite{CS09b}. In the present paper,
we focus on a simpler approach, based upon the fixed-energy MPMSA.

Note also that the decay bounds \ref{eq:thm.Main.DL.1}--\ref{eq:thm.Main.DL.2}
have an unusual form as compared to their more traditional counterparts from
the single-particle MSA: constants $C(\hBx_j, \om)$, $C(\Bx)$ depend upon the
multi-particle configuration $\Bx$ (actually, upon the diameter of its support,
$\diam( \{x_1, \ldots, x_N\})$). This feature is present in the first
papers on multi-particle localization \cite{CS09b}, \cite{AW09a}; we discuss it below in
more detail.

As was already said, in order to obtain more optimal localization bounds, viz.:
exponential spectral localization
(as in \cite{CS09b}) and uniform estimates on the decay of eigenfunctions and
their correlators, we make use of a stronger assumption \Wtwo.

\begin{theorem}\label{thm:Main.DL.Wtwo}
Fix any integer ${\hN}\ge 2$.
Under the assumptions {\rm \Wtwo} and {\rm\Uone}, for any  $N\in[1,{\hN}]$ there exists $g_0(N)<+\infty$ such that
for all $|g|\ge g_0(N)$ and some $m=m(g,N)\ge Const(N) \ln |g|>0$:
\par\noindent
{\rm{\textbf{(A)}}}
with probability one, the random operator $\BH(\om)=\BH_0 + g\BV(\om)+\BU$
has pure point spectrum and all
eigenfunctions $\BPsi_j(\om)$ decay exponentially fast at infinity:  for each $\BPsi_j$,
some $\widehat{\Bx}_j(\om)\in\bcZN$, $C_j(\om)\in(0,+\infty)$ and all $\Bx\in\bcZN$,
\be\label{eq:thm.Main.DL.1.Wtwo}
| \BPsi_j(\Bx,\om) | \le C_j(\om) \,\eu^{-m \brho_{\rS}(\Bx, \hBx_j)} ;
\ee
\par\noindent
{\rm{\textbf{(B)}}}
for all points $\Bx,\By$ and some $a,c, C\in(0,+\infty)$,
\be\label{eq:thm.Main.DL.2.Wtwo}
\esm{ \sup_{f\in\csB_1} \big| \langle \one_{\By}\, | \,  f(\BH(\omega)) \, | \, \one_{\Bx}\rangle\big| }
\le C \,\eu^{ -a\ln^{1+c} \brho_\rS(\Bx,\By) }.
\ee
\end{theorem}

\vskip4mm
It is readily seen that, given a point $\Bu\in\bcZN$ (e.g., $\Bu=\hBx_j$),
$$
\brho(\Bx, \Bu) \ge \brho\big(\Bx, \{\pi(\Bu), \, \pi\in\fS_N\} \big)
\tto{ \brho(\Bx,\Bu) \to +\infty } +\infty,
$$
\vskip2mm\noindent
so that the assertions (A) and (B) of Theorem \ref{thm:Main.DL.Wtwo} imply
their counterparts from Theorem \ref{thm:Main.DL.Wone} (under a more restrictive
hypothesis \Wtwo). An important point is that Theorem \ref{thm:Main.DL.Wtwo}
provides a more precise and explicit information on the decay
properties of the EFs and their correlators. Moreover, once formulated
in a physically relevant language of \emph{indistinguishable} particles
(bosons or fermions), Theorem \ref{thm:Main.DL.Wtwo} operates with
a \emph{natural distance} $\brho_{\rS}$ between particle configurations. Possible
lack of decay in terms of the
non-symmetrized distance $\brho$ is due to the fact that any $\pm$-symmetric
eigenfunction $\BPsi(\Bx)$ takes the values of the same magnitude
along an orbit $\csO(\Bx):=\{\pi(\Bx), \,\pi\in\fS_N\}$.

Working with restrictions of the Hamiltonian $\BH^{(N)}(\om)$ to the subspaces
of symmetric or antisymmetric functions on $\cZ^N$ would be simpler on
the $N$-th symmetric power of the 1-particle (physical) configuration space $\cZ$.
In the framework of the variable-energy scaling analysis, such an approach has been
used in our earlier work \cite{C11b}. Here, for the sake of brevity, we prefer
to work with distinguishable particles, i.e., with Hamiltonians $\BH^{(N)}(\om)$
in the entire Hilbert space $\ell^2(\cZ^N)$. This also leads, formally,
to more general results. Nevertheless, it is to be stressed again that
only the language of indistinguishable particles is physically relevant.

Finally, note that adaptations of the variable-energy
MPMSA to localization
near the lower edge of  the spectrum
(without the assumption of strong disorder) and to weakly interacting
systems ("sufficiently" localized without interaction) have been obtained
by Ekanga \cite{Ek11,Ek12}, in the case where $\cZ=\DZ^d$,
under the assumption of log-H\"{o}lder continuity
of the PDF $F_V$ of the external random potential.

\vskip4mm

Theorems \ref{thm:Main.DL.Wone} and \ref{thm:Main.DL.Wtwo} give the final results
of the multi-particle localization analysis. However, the novelty
of the present work resides essentially in Theorems \ref{thm:from.VEMPMSA.to.DL.W1}
and \ref{thm:from.VEMPMSA.to.DL.W2}, deriving the spectral and dynamical localization
from the results of the fixed-energy analysis of resolvents.

\section{Subharmonicity on graphs}
\label{sec:SH}

The results of this section apply to arbitrary graphs, including
$\cZ^N$, $N\ge 2$.

\begin{definition}\label{def:subh}
Let $\cG$ be a finite connected graph, $L\ge \ell\ge 0$ two integers and $q\in(0,1)$.
A function $f:\cG\to\DR_+$ is called $(\ell,q)$-subharmonic in a ball $\ball_L(u)\subsetneq \cG$
if for any ball $\ball_\ell(x)\subseteq\ball_L(u)$ one has
\be\label{eq:lqsubh}
f(x) \le q \max_{y\in\ball_{\ell + 1}(x)} f(y).
\ee
\end{definition}

We will often use the notation $\cM(f,\Lam) := \max_{x\in\Lam}| f(x)|$.
\begin{lemma}\label{lem:subh.1}
If a function $f:\cG\to\DR_+$ defined on a finite connected graph $\cG$ is $(\ell,q)$-subharmonic
in a ball $\ball_L(x)\subsetneq\cG$, with $L\ge \ell\ge 0$, then
\be\label{eq:lem.subh.1}
f(x) \le q^{ \left\lfloor \frac{L+1}{\ell+1} \right\rfloor} \cM(f,\cG)
     \le q^{ \frac{L - \ell }{\ell+1} } \cM(f,\cG).
\ee
\end{lemma}

In fact, the factor $\cM(f,\cG)$ in the RHS of \eqref{eq:lem.subh.1} can be replaced
by $\cM(f,\ball_{L+1}(x))$.

\proof
See \cite{C12a}.
\qedhere
\vskip4mm

\begin{lemma}\label{lem:GF.is.SH.1}
Consider a finite connected graph $\cG$ and a ball $\ball_L(u)\subsetneq \cG$, with
$L\ge \ell\ge 0$. Fix numbers $E\in\DR$, $m>0$ and suppose that all balls $\ball_\ell(x)$
inside $\ball_L(u)$ are \EmNS.
Then $\forall$ $y\in\cG\setminus\ball_L(u)$ the function
$$
f: x \mapsto |G_{\cG}(x,y;E)|
$$
is $(\ell,q)$-subharmonic in $\ball_L(u)$ with $q = \eu^{-\gamma(m,\ell)\ell}$.
\end{lemma}

\proof The claim follows directly from the Definition \ref{def:subh}.
\qedhere

\vskip 2mm

Lemma \ref{lem:subh.1} suffices to assess the Green functions in a ball $\ball_L(u)$ which does
not contain any singular $\ell$-ball, but to analyze the situation where $\ball_L(u)$
does not contain any \emph{pair} of disjoint
singular $\ell$-balls, one needs the following extension of Lemma \ref{lem:subh.1},
inspired by the proof of Theorem 1 in \cite{Sp88}: approaching a single "bad" ball separately
from the points $x$ and $y$.

\begin{lemma}\label{lem:subh.2}
Let $\cG$ be a finite connected graph, and $f:\cG\times\cG\to\DR_+$, $f:(x,y)\mapsto f(x,y)$, be a function which is separately $(\ell,q)$-subharmonic in $x\in\ball_{r'}(u')\subset \cG$ and in
$y\in\ball_{r''}(u'')\subset \cG$, with $r',r''\ge \ell\ge 0$ and $\rd(u', u'')\ge r'+r''+2$. Then
\be\label{eq:lem.subh.2}
f(u',u'') \le q^{ \left\lfloor \frac{r'+1}{\ell+1} \right\rfloor
                 + \left\lfloor \frac{r''+1}{\ell+1} \right\rfloor } \cM(f,\cG\times\cG)
           \le q^{ \frac{r'+r'' - 2\ell}{\ell+1} } \cM(f,\cG\times\cG).
\ee
\end{lemma}

\proof
See \cite{C12a}.
\qedhere

\section{Eigenvalue concentration bounds}
\label{sec:EVC}

\subsection{One-volume EVC bounds}
The first eigenvalue concentration (EVC) bound required for our scheme is
a direct analog of the Wegner estimate \cite{W81} for multi-particle operators. It can be proved under the assumption \Wone.
\begin{theorem}
\label{thm:W1}
Let $V: \cZ\times \Omega \to \DR$ be a random field satisfying the hypothesis
{\rm\Wone}.
Then for any $\beta'\in(0,\beta)$ and large $L_0$, the following bound holds true:
\be\label{eq:thm.W1}
\forall\, \Bx\in\cZ^N\;\; \sup_{E\in\DR}  \;
\pr{ \text{ $\bball_{L}(\Bx)$ is {\rm \ER} } } \le e^{-L^{\beta'} }
\ee
\end{theorem}

With $\beta = 1/2$, it suffices to set $\beta'=1/4$.

Naturally, Theorem \ref{thm:W1} is based on a probabilistic bound on the events of the
form $\{\om:\,\dist(E, \Sigma(\BH_{\bball_L(\Bx)})) < s\}$, with $s = \eu^{-L^\beta}$.
An optimal Wegner-type bound, for an IID random potential $V$ with Lipshitz-continuous
marginl PDF $F_V$, would have the form
\be\label{eq:Wegner.usual}
\pr{\om:\,\dist(E, \Sigma(\BH_{\bball_L(\Bx)})) < s} \le \Const |\bball_L(\Bx)| \, s.
\ee
Indeed, such a (multi-particle)
bound was proved by the author and Yuri Suhov
for IID random potentials with PDF analytic in a
strip $\{E\in\DC:\, |\Im E| < \eps\}$, for some $\eps>0$ (hence, having
an analytic probability density).  This covers the so-called $\alpha$-stable laws
indexed by a real parameter $\alpha\in[1,2]$ (not to be confused with the scaling exponent
$\alpha$ appearing in the recursion $L_{k+1} = [L_k^\alpha]$), where $\alpha=2$ corresponds
to the Gaussian law and $\alpha=1$ to the Cauchy law. The proof is based on Molchanov's
rigorous path integral formula for the unitary propagators $\eu^{-\ii t H}$ (for finite-difference
operators $H$) and follows essentially the same path as in Carmona's proof of analyticity
of the Density of States (DoS) for tight-binding Anderson Hamiltonians; cf. \cite{CL90}.
The same method applies to the two-volume bounds given in Theorem \ref{thm:W2}
below, again for analytic PDF $F_V$.

Kirsch\footnote{See also the preprint \texttt{arXiv:math-ph/0704.2664} (2007).} \cite{K07} proved an optimal multi-particle Wegner estimate of the
form \eqref{eq:Wegner.usual} for uniformly Lipshitz continuous marginal
probability distributions. The proof of Theorem \ref{thm:W1}
in \cite{CS08}, for possibly singular (but uniformly continuous) marginal
probability distributions is based on Stollmann's lemma from \cite{St00} which gives rise to a 
non-optimal volume dependence; the latter, however, is absorbed in the fractional exponent
and is more than sufficient for applications to the multi-particle MSA. Stollmann's lemma itself
is very general and not sensitive to the regularity properties of the marginal distribution, but
is effective (suitable for MSA applications)
only for marginal PDFs featuring at least log-H\"{o}lder continuity, with suitable
parameters.

A one-volume EVC multi-particle bound for continuous systems (Anderson Hamiltonians with
an alloy-type random potential), with optimal
volume dependence, has been reported by Hislop and Klopp \cite{HK12}. A variant
of such bound (again, for alloy-type potentials in $\DR^d$) based on Stollmann's lemma,
with a non-optimal volume dependence, was proven earlier in \cite{BCSS09}
and used in \cite{CBS11}, in the proof of multi-particle dynamical localization
for Anderson-type Hamiltonians in $\DR^d$.

\subsection{"Separability" in the multi-particle configuration space}

All above mentioned one-volume EVC bounds suffice for the proof of the a.s.
exponential decay (hence, square summability) of \textbf{\emph{multi-particle}}
Green functions.
This point was not clear earlier, since Simon--Wolf method from \cite{SW86} does not
apply to multi-particle Hamiltonians. However, both the variable-energy MPMSA from
\cite{CS09b} and the derivation of the VEMPMSA bounds from their fixed-energy counterparts
(presented here) require two-volume EVC bounds which are more sophisticated
for multi-particle systems than in the conventional, single-particle localization
theory. It turns out that the positions of the balls play an important role. For this reason, following \cite{CS09b}, we introduce 

\begin{definition}
A ball $\bball^{(N)}_L(\Bx)$  is separable from a ball $\bball^{(N)}_L(\By)$
if there exists a decomposition
$
\bball^{(N)}_L(\Bx) = \bball^{(n')}_L(\Bx') \times \bball^{(n'')}_L(\Bx'')
$
such that (cf. Eqn \eqref{eq:def.Pi.ball})
\be\label{eq:def.sep}
 \Pi \bball^{(n')}_L(\Bx') \cap \Pi \bball^{(N)}_L(\By) = \varnothing.
\ee
A pair $(\bball^{(N)}_L(\Bx), \bball^{(N)}_L(\By))$ is separable
if one of the balls is separable from the other.
\end{definition}

Pictorially, \eqref{eq:def.sep} says that there is a sub-sample of the potential
which determines the operator $\BH_{\bball_L(\By)}$ completely, while
$\BH_{\bball_L(\Bx)}$ (more precisely, its component relative to a sub-configuration
$\Bx'$) has a non-degenerate (and controllable) probability distribution, conditional
on $\BH_{\bball_L(\By)}$.

\begin{lemma}\label{lem:3NL.WS}
For any $\Bx\in\cZ^N$ there is $R(\Bx)$ such that any ball $\bball^{(N)}_L(\By)$
with $\brho(\Bx, \By) \ge R(\Bx)$ is separable from $\bball^{(N)}_L(\Bx)$.
\end{lemma}
\proof
Let $r = \diam( \Pi \Bx)$ and set $R=R(\Bx) = r + 3NL$. Consider a ball
$\bball^{(N)}_L(\By)$ with $\brho(\Bx, \By) > R$. By definition of the max-distance,
there is at least one particle position $y_{i_0}\in \Pi \By$ such that
$\rd(y_{i_0}, x_{i_0}) \ge R$. Consider the maximal connected component
$\BLam_\By:=\cup_{i\in \cJ} \ball_{2L}(y_i)$ of the union $\cup_i \ball_{2L}(y_i)$ containing
$y_{i_0}$; its diameter is bounded by $2NL$, and by triangle inequality,
$$
\dist\left( \BLam_\By, \Pi \bball_L(\Bx) \right)
\ge R - (\diam \,\Pi\Bx + 2L) - \diam\, \BLam_\By > 0.
$$
Taking as the subconfiguration $\By'$ the union $\cup_{i\in\cJ} \{y_i\}$,
we conclude that $\bball^{(N)}_L(\By)$  is indeed separable from $\bball^{(N)}_L(\Bx)$.
\qedhere

\begin{theorem}[Cf. Lemma 2 in \cite{CS09b}]\label{thm:W2}
Under the hypothesis {\rm\Wone},
the following bound holds true for any pair of separable balls $\bball_{L}(\Bx)$,
$\bball_{L}(\By)$ and all $s\in[0,1]$:
\be\label{eq:thm.W2}
 \pr{ \dist\Big( \Sigma\big(\BH_{\bball_{L}(\Bx)}\big),
                  \Sigma\big(\BH_{\bball_{L}(\By)}\big) \Big) \le s }
\le \Const\, L^{(2N+1)d} s^\delta.
\ee
\end{theorem}

Lemma 2  in \cite{CS09b} is formally stated for operators on a lattice
$(\DZ^d)^N$, but its proof, based on a very general Stollmann's lemma on monotone functions
\cite{St00,St01}, applies, with minor notational modifications, also to Hamiltonians
on graphs, provided that the extrenal random potential
field $V:\cZ\times\Om\to\DR$ is IID with H\"{o}lder-continuous marginal distribution.

The dependence of the upper bound on the EF correlators in Theorem \ref{thm:Main.DL.Wone}
upon the positions $\hBx_j$, $\Bx$ is explained by the nature of the two-volume EVC bound
in Theorem \ref{thm:W2}: under the hypothesis \Wone, 
"tunneling" between quantum states $\Bx$, $\By$ is ruled
out with high probability only if the balls $\bball_L(\Bx)$, $\bball_L(\By)$, with
a suitable $L = O(\brho(\Bx,\By))$, are separable.

\subsection{"Weak separability" at large distances}

\begin{definition}\label{def:ws}
A  ball $\bball^{(N)}_L(\Bx)\subset\bcZN$ is weakly separable from $\bball^{(N)}_L(\By)$ if
there exists a ball $\Lam\subset \cZ$ in the 1-particle configuration space,
of diameter $R \le 2NL$,  and  subsets       $\cJ_1, \cJ_2\subset [1,N]\cap\DZ$  such that
$|\cJ_1| > |\cJ_2|$ (possibly,  $\cJ_2=\varnothing$) and
\begin{equation}\label{eq:cond.WS}
\left\{\begin{array}{l}
 \PPi_{\cJ_1} \bball^{(N)}_L(\Bx) \cup \PPi_{\cJ_2} \bball^{(N)}_L(\By) \;  \subseteq \Lam,
\\
\PPi_{\cJ^c_1} \bball^{(N)}_L(\Bx) \cap \Lam = \varnothing,
\\
\PPi_{\cJ^c_2} \bball^{(N)}_L(\By) \cap \Lam = \varnothing.
\end{array}
\right.
\end{equation}
A  pair of balls $(\bball^{(N)}_L(\Bx), \bball^{(N)}_L(\By))$ is weakly separable if at least one of the balls is weakly separable from the other.
\end{definition}

The physical meaning of the weak separability is that in a certain region of the one-particle configuration space, there are more particles from configuration $\Bx$ than from $\By$. As a result, some local fluctuations of the random potential $V(\cdot;\omega)$ have a stronger influence on EVs relative to  $\bball_{L}(\Bx)$ than on EVs relative to  $\bball_{L}(\By)$. It is easy to see that this condition is indeed weaker than the separability.

\begin{lemma}[Cf. Lemma 2.3 in \cite{C10}]\label{lem:dist.are.WS}
If $\brhoS(\Bx, \By)> 3NL$, then balls $\bball^{(N)}_L(\Bx)$ and $\bball^{(N)}_L(\By)$  are weakly separable.
\end{lemma}

The lower bound by $3NL$ is not sharp, but slightly less cumbersome than the one established in \cite{C10}; the actual value of the constant (in front of $NL$) is irrelevant.
The proof is given in \cite{C10}, formally, applies to the case where
$\cZ = \DZ^d$, but the extension to more general graphs is straightforward.
The main idea can be summarized as follows: two balls of radius $L$ with centers $\Bx, \By$
\emph{are not} weakly separable iff all particles
from $\Bx$ occupy approximately the same positions as the particles from $\By$
(up to a permutation: this is why the symmetrized distance $\brhoS$ is used). Specifically,
there has to be a permutation $\pi\in\fS_N$ such that $\forall\,i\in[1,N]$,
$\rd(x_i, y_{\pi(i)}) = O(L)$, yielding $\brhoS(\Bx, \By) = O(NL)$.

We will call a pair of balls $\bball^{(N)}_L(\Bx), \bball^{(N)}_L(\By)$
\emph{distant} iff $\brhoS(\Bx, \By) \ge 3NL$.

\begin{theorem}[Cf. Lemma 3.1 in \cite{C10}]\label{thm:EVC.PCT.Wtwo}
Under the hypothesis {\rm \Wtwo}, for any pair of weakly separable (e.g., distant)
balls $\bball_{L}(\Bx)$, $\bball_{L}(\By)$  the following bound holds true:
$$
\pr{ \dist\big(\Sigma(\BH_{\bball_{L}(\Bx)}),\Sigma(\BH_{\bball_{L}(\By)}) \big) \le s }
     \le  h_L(2s),
$$
with $h_L(s)$ defined in \eqref{eq:h.L.weak.sep} and constants $C,C',A, A',B,B'$
defined in {\rm\Wtwo}:
\be\label{eq:h.L.weak.sep.again}
h_L(s) = |\bball_{L}(\Bx)| \cdot |\bball_{L}(\By)| C L^A s^B + C' L^{A'} s^{B'}.
\ee
\end{theorem}

The proof given in \cite{C10} in the case of a periodic lattice $\cZ = \DZ^d$
extends with no difficulty to more general graphs.

\section{Fixed-energy multi-particle scale induction}
\label{sec:FEMSA}

It is to be emphasized that the entire fixed-energy scaling procedure, for any number
$N\ge 2$ particles, does not reveal the difficulty, mentioned in the Introduction and related
to the "structural" resonances. As a result,
the exponential decay (hence, square-summability) of Green functions at a fixed
energy can be established under a weaker hypothesis of H\"{o}lder (or log-H\"{o}lder)
continuity of the marginal PDF $F_V$ of the random potential field
$V:\cZ\times\Om\to\DR$.

\subsection{Scaling of Green functions in absence of tunneling}

\begin{definition}
A ball $\bball_{L_{k+1}}(\Bu)$ is called $E$-tunneling ({\rm\ET}) if it contains two disjoint
\EmS balls of radius $L_k$, and $E$-non-tunneling ({\rm\ENT}), otherwise.
\end{definition}

\begin{lemma}\label{lem:NR.NT.implies.NS}
If a ball $\bball_{L_{k+1}}(\Bu)$ is {\rm\ENR} and {\rm\ENT}, then it is
{\rm\EmNS}.
\end{lemma}

\proof See the proof of Lemma 5.1 from \cite{C12b}. Its statement is deterministic
and not specific to the analytic nature of the potential, thus applies
to DSO on graphs with single- and multi-particle structure of the potential.
\qedhere

\subsection{Main inductive bound}

Given a number $\kappa> \frac{2\alpha}{2-\alpha}Nd$, pick any
$0 < \theta < \frac{2\alpha}{2-\alpha} - \frac{2Nd}{\kappa}$ and introduce the following double sequence:
\be\label{eq:def.p.n}
P(N,k) = P(N,k; \hN, \kappa, \theta) = 3^{{\hN}-N} \kappa(1+\theta)^k, \; N=1, \ldots, {\hN}; \; k\ge 0,
\ee
and a statement, or property, depending upon integer parameters $N\in[1,\hN]$, $k\ge 1$,
and an interval $I\subseteq \DR$,
which we are going to prove by induction:

\par
\medskip\noindent
\SSI{N,k}: \qquad $\forall\, E\in I$\; $\forall\, n\in[1,N]$ $\;\;\forall$ $\Bx\in\bcZ^n$

\be\label{eq:nt}
\pr{ \text{$\ball^{(n)}_{L_{k}}(\Bx)$ is \EmS } }
\le  L_{k}^{ -3^{\hN - n} \kappa } = L_{k}^{ -P(n,k) }.
\ee

\subsection{Initial scale bounds}

First, consider the case of strongly disordered systems.

\begin{lemma}
For any $L_0\ge 2$, $m\ge 1$, $\hN\ge 1$ and $N\in[1,\hN]$ there is
$g^* = g^*(m,p,n,L_0) < \infty$ such that if $|g|\ge g^*$, then
the property \SSI{N,0} holds true.
\end{lemma}

The proof is similar to its well-known 1-particle counterpart (cf., e.g.,
\cite{DK89}) and will be omitted. It also follows easily from the Wegner-type estimate
(cf. Theorem \ref{thm:W1}) combined with the Combes--Thomas estimate.

For energies close to the lower edge of the spectrum, an initial scale
bound suitable for the purposes of the MPMSA has been proved by Ekanga \cite{Ek11,Ek12}.

In the framework of multi-particle systems with interaction, a third scenario
appears where the localization can be proved, merely by an adaptation of the
initial scale estimate: systems localized without interaction and perturbed
by a sufficiently weak interaction (e.g., of short range). Such scenario has
been first analyzed by Aizenman and Warzel \cite{AW09a} with the help
of the MPFMM. Ekanga \cite{Ek12} adapted the method of \cite{CS09b}
to weakly interacting systems. In particular, he proved that $N$-particle
systems in $\DZ^1$, with two-parameter Hamiltonian
$$
\BH = -\BDelta + g\BV(\Bx;\om) + h\BU(\Bx)
$$
are localized for any nonzero $g\in\DR$ and all sufficiently
small $|h|\le h^*(|g|)$. Note, however, that the random potential
is required to have log-H\"{o}lder continuous marginal PDF $F_V$
(with suitable parameters), so singular (e.g., Bernoulli) potentials are not allowed
in \cite{Ek12}. Indeed, the proof follows the MSA scheme (which is perturbative,
by its nature), for the specifically one-dimensional methods
based on Furstenberg theory do not apply even to 2-particle Hamiltonians.
The same difficulty is encountered in \cite{AW09a,AW09b} where
the fractional moment method is also perturbative.

\begin{lemma}[Cf.  \cite{Ek11,Ek12}]
For any $L_0\ge 2$, $m\ge 1$, $\hN\ge 1$ and $N\in[1,\hN]$ there is
$g^* = g^*(m,p,n,L_0) < \infty$ such that if $|g|\ge g^*$, then
the property \SSI{N,0} holds true.
\end{lemma}

\proof
(We only sketch the proof.)
Owing to the positivity of interaction $U$, of the external potential $V$
and of the graph Laplacian $-\Delta$, one has
$$
\BH^{(N)} \ge H^{(1)}\otimes \one^{(N-1)}.
$$
Therefore, a lower bound for the spectrum of $\BH^{(N)}$
is provided, e.g., by the lowest eigenvalue of the single-particle Hamiltonian.
For the latter, it is well-known and is due to the Lifshitz tails
asymptotics.
\qedhere

\vskip2mm

The case of the random potential bounded from below by an arbitrary quantity
$E_*>-\infty$ can be treated similarly, replacing $V(x;\om)$ by $V(x;\om) - E_0$
and $\BH^{(N)}$ by $\BH^{(N)} - NE_0\one$. In the case where $E_0 = -\infty$ (e.g., Gaussian
potentials) the proof is even easier and follows  virtually
the same path as for strongly disordered systems.

For brevity, below we treat \textbf{only the case of strong disorder}; an adaptation
to the band-edge localization merely requires the control of the "mass" $m>0$
and simple modifications of scaling formulae.

\subsection{Green functions in decoupled systems}

\begin{definition}\label{def:PI}
An $N$-particle ball $\boxx_L(\Bu)$ is called partially interactive (\PI) if
$\diam( \Pi \Bu) > 11 N L$, and fully interactive (\FI), otherwise.
\end{definition}

If the interaction has finite range $r_0$ and $L \ge 8r_0$,  a polydisk
$\bball_L(\Bu)$ can be represented as follows:
\be\label{eq:balls.decomposable}
\bball_L(\Bu) = \bball^{(N)}_L(\Bu)
= \bball^{(n')}_L(\Bu') \times \bball^{(n'')}_L(\Bu'')
\ee
where $n'+n''=N$ and the subconfigurations
$\Bu=(\Bu',\Bu'')$, $\Bu'\in\cX^{n'}$, $\Bu''\in\cX^{n''}$ satisfy the condition
$$
\rho\left(\PPi \bball^{(n')}_L(\Bu'), \PPi \bball^{(n'')}_L(\Bu'') \right)
> r_0.
$$
As a result, the operator $\BH^{(N)}_{\bball_L(\Bu)}$ in
$\bball_L(\Bu)$
is algebraically decomposable  in the following way:
\be\label{eq:Ham.decomposable}
\BH^{(N)}_{\bball_L(\Bu)} = \BH^{(n')}_{\bball_L(\Bu')}\otimes \one^{(n'')}
+ \one^{(n')} \otimes \, \BH^{(n'')}_{\bball_L(\Bu'')}
\ee
since, for any $\Bx=(\Bx', \Bx'')\in\bball^{(N)}_L(\Bu)$, the interaction  between the sub-con\-figurations $\Bx'$ and $\Bx''$ vanishes. Such a decomposition, if it exists,
may be non-unique. We will assume that one such decomposition (referred to as
the canonical one) is associated in some way with every $N$-particle \PI ball.

\begin{lemma}\label{lem:PITRONS}
Assume that the interaction $\BU$ satisfies the condition {\rm\Uone}. Fix an energy $E\in\DR$.
Consider a {\rm PI} $N$-particle ball with canonical decomposition
$\bball^{(N)}_{L_k}(\Bu) = \bball^{(n')}_{L_k}(\Bu') \times \bball^{(n'')}_L(\Bu'')$
and a sample of the potential $V$ such that
\begin{enumerate}[{\rm(a)}]
  \item $\bball^{(N)}_{L_k}(\Bu)$ is {\rm\ENR}.

  \item $\forall\,\lam''\in\Sigma(\BH_{\bball^{(n'')}_{L_k}(\Bu'')})$ the ball $\bball^{(n')}_{L_{k}}(\Bu')$  is $(E-\lam'',m)${\rm-NS};

  \item $\forall\,\lam'\in\Sigma(\BH_{\bball^{(n')}_{L_k}(\Bu')})$ the ball $\bball^{(n'')}_{L_{k}}(\Bu'')$  is $(E-\lam',m)${\rm-NS}.
\end{enumerate}
Then the ball $\bball^{(N)}_{L_k}(\Bu)$ is {\rm\EmNS}.
\end{lemma}
\proof See subsection \ref{ssec:proof.PITRONS}. \qedhere

\begin{corollary}\label{cor:prob.PI.S}
Assume the property {\rm\SSI{N,k}} and let $L_0\ge 4C_d^N$.
Then for all $E\in I$ and any {\rm PI} ball of the form
$\bball^{(N)}_{L_k}(\Bx)\subset\cZ^N$,
\be\label{eq:lem.PI.T}
\pr{\text{ $\ball^{(N)}_{L_k}(\Bx)$  is   {\rm\EmS} }}
\le \frac{1}{4}  C_d^{-2N} L_k^{ -\frac{11}{4}P(N,k)  }
\ee
\end{corollary}

\proof
Denote by $\cS_k$ the event in the LHS of \eqref{eq:lem.PI.T}.
Consider the canonical decomposition $\bball^{(N)}_{L_{k}}(\Bx)=\bball^{(n')}_{L_{k}}(\Bx')\times\bball^{(n'')}_{L_{k}}(\Bx'')$.
By virtue of Lemma \ref{lem:NR.NT.implies.NS},
\be\label{eq:proof.lem.PI.T}
\bal
\pr{\cS_k }
< \pr{\text{ $\ball^{(N)}_{L_k}(\Bx)$  is   \ER }}
+ \pr{\text{ $\ball^{(N)}_{L_k}(\Bx)$  is  \ENR and \EmS }}.
\eal
\ee
The first term in the RHS is bounded with the help of Theorem \ref{thm:W1}, so we focus on the second term.
Apply Lemma \ref{lem:PITRONS}: since the option (a) is ruled out, it remains to assess the probability of events listed in options (b) and (c). Consider the former:
$$
\bal
\pr{\cS' } &\equiv
 \pr{ \text{ $\exists\,\lam''\in\Sigma(\BH_{\bball^{(n'')}_{L_k}(\Bx'')})$\,:\;  $\bball^{(n')}_{L_{k}}(\Bx')$  is $(E-\lam'',m)${\rm-S}} }
\\
& =\esm{ \pr{
 \exists\,\lam''\in\Sigma(\BH_{\bball^{(n'')}_{L_k}(\Bx'')}):\, \bball^{(n')}_{L_{k}}(\Bx')
   \text{ is } (E-\lam'',m){\rm-S}} \,\Big|\, \fF''}
\eal
$$
where $ \fF''$ is the sigma-algebra generated by
the values of the random potential in $\bball^{(n'')}_{L_k}(\Bx'')$.
By definition of the canonical decomposition of a PI ball,
\be\label{eq:dist.proj.big}
 \PPi \,\bball^{(n')}_{L_{k}}(\Bx') \cap  \PPi \,\bball^{(n'')}_{L_{k}}(\Bx'') =\varnothing,
\ee
and since the random field $V$ is IID, for any $E''\in \DR$, including
$E-\lam''$, one has
\be\label{eq:use.mixing}
{\rm(a.s.)}\quad
\pr{\text{ $\bball^{(n')}_{L_{k}}(\Bx')$  is $(E'',m)${\rm-S}} \,\big|\, \fF''}
= \pr{\text{ $\bball^{(n')}_{L_{k}}(\Bx')$  is $(E'',m)${\rm-S}} }.
\ee
On the other hand, by the assumed property \SSI{N-1,k}, for $n'\le N-1$,
\be\label{eq:proj.S.prob}
\pr{ \text{  $\bball^{(n')}_{L_{k}}(\Bx')$  is $(E'',m)${\rm-S}} }
\le C_d^{-2N} L_k^{ - P(N-1,k)} = C_d^{-2N} L_k^{ - 3 P(N,k)}.
\ee
Therefore, we obtain, with $L_0\ge 4C_d^N$,
\be\label{eq:proof.lem.PI.T.2}
\bal
\pr{\cS' }
& \le |\bball^{(n'')}_{L_k}(\Bx'')|\, \sup_{E''\in\DR}
\pr{ \text{$\bball^{(n')}_{L_{k}}(\Bx')$  is $(E'',m)${\rm-S}} }
\\
&\le \frac{C_d^N}{C_d^{2N}} L_k^{Nd}\, L_k^{ - 3 P(N,k)}
\le \frac{1}{4} C_d^{-2N} L_k^{ -\frac{11}{4}P(N,k) - (\frac{1}{4} P(N,k) - Nd-1)}
\\
&\le \frac{1}{4}  C_d^{-2N} L_k^{ -\frac{11}{4}P(N,k)  }
\eal
\ee
since $p > 6Nd$, so $\frac{1}{4} P(N,k) > \frac{3}{2}Nd \ge Nd+1$ (for $N\ge 2$).
Similarly,
\be\label{eq:proof.lem.PI.T.3}
\pr{\text{ $\exists\,\lam'\in\Sigma(\BH_{\bball^{(n')}_{L_k}(\Bx')})$\,:\; $\bball^{(n'')}_{L_{k}}(\Bx'')$  is $(E-\lam',m)${\rm-S}}}
\le \frac{1}{4}  C_d^{-2N} L_k^{ -\frac{11}{4}P(N,k)  }.
\ee
Collecting \eqref{eq:thm.W1}, \eqref{eq:proof.lem.PI.T},
\eqref{eq:proof.lem.PI.T.2} and
\eqref{eq:proof.lem.PI.T.3}, the assertion follows.
\qedhere

\subsection{Scale induction}

Introduce the following notations: with $k\ge 0$,
$$
\bal
\rP_k &= \sup_{\Bx\in\bcZN} \pr{ \bball_{L_k}(\Bx) \text{ is \EmS}},
\\
\rQ_{k+1} &= 4 \sup_{\Bx\in\bcZN} \pr{ \bball_{L_{k+1}}(\Bx) \text{ is \ER}},
\\
\rS_{k+1} &= \sup_{\Bx\in\bcZN} \pr{ \bball_{L_k}(\Bx)
\text{ contains a PI \EmS ball of radius $L_{k}$}}.
\eal
$$
For future use, note that by Corollary \ref{cor:prob.PI.S}, with $\kappa >6Nd$ and
$0<\theta < 1/6$,
\be\label{eq:prob.S.k+1}
\bal
\rS_{k+1} &\le
\frac{1}{2 C_d^{2N}}   L_k^{ -\frac{11}{4}P(N,k)  } \cdot C^N_d L_{k+1}^{Nd}
\le \frac{1}{4 C_d^{2N}} L_{k+1}^{ -\frac{11}{4} \cdot \frac{2}{3} P(N,k)  -Nd }
\\
&\le \frac{1}{4 C_d^{2N}} L_{k+1}^{ -\frac{11}{6(1+\theta)} \kappa(1+\theta)^{k+1}   -Nd}
\le \frac{1}{4 C_d^{2N}} L_{k+1}^{ -\frac{11}{6(1+\frac{1}{6})} \kappa(1+\theta)^{k+1}  -Nd }
\\
&\le \frac{1}{4 C_d^{2N}}\, L_{k+1}^{-\frac{11}{7}  \kappa(1+\theta)^{k+1} -Nd}
\le \frac{1}{4 C_d^{2N}}\, L_{k+1}^{-\kappa(1+\theta)^{k+1}}.
\eal
\ee

\begin{theorem}\label{thm:MSA.main.fixed}
Let $\alpha\in(1,2)$, $\kappa>\frac{2Nd\alpha}{2-\alpha}$ and
$\theta\in(0, \frac{2-\alpha}{\alpha} - \frac{2Nd}{\kappa})$;
assume the condition {\rm\Wone}.
If there is an integer $L_0\ge 4 C_d^N$ such that
{\rm\SSI{N,0}} is fulfilled and
$\rQ_0 \le C_d^{-2N} L_{0}^{-\kappa}$,
then {\rm\SSI{N,k}} holds true for all $k\ge 0$.
\end{theorem}

\proof
It suffices to derive \SSI{N,k+1} from \SSI{N,k}, so assume the latter.
By virtue of Lemma \ref{lem:NR.NT.implies.NS}, if a ball $\ball_{L_{k+1}}(u)$ is \EmS,
then it is either \ER or \ET. By Eqn \eqref{eq:prob.S.k+1}, the probability
to have at least one PI
\EmS ball of radius $L_k$ inside $\ball_{L_{k+1}}(u)$ obeys
$\rS_{k+1}\le \frac{1}{4 C_d^{2N}}\, L_{k+1}^{-\kappa(1+\theta)^{k+1}}$.

Further, there are $<\half C^{2N}_d L_{k+1}^{2Nd}$
pairs of disjoint  $L_k$-balls in $\ball_{L_{k+1}}(u)$, thus
$$
\rP_{k+1} \le \half C_d^{2N} L_{k+1}^{2Nd} \rP_k^2 + \frac{1}{4} \rQ_{k+1} + \rS_{k+1}.
$$
By Theorem \ref{thm:W1},
$\rQ_{k+1} \le Const\, L_{k+1}^{Nd}  \eu^{-L_{k+1}^{\beta}}$.
The function
$$
\bal
f: L &\mapsto \ln \left( \const L^{-\kappa}\right)
-\ln \left(\const\, L^d  \eu^{-L^{\beta}} \right)
%
= L^{\beta} - \const \ln L
\eal
$$
on $[1, +\infty)$ is either non-negative or admits a unique zero. In either case, the assumption
$\rQ_0 \le C_{d}^{-2N} L_{0}^{-\kappa}$ implies $\rQ_{k+1} \le C_d^{-2N} L_{k+1}^{-\kappa}$
for all $k\ge 0$, so
$$
\frac{1}{4}\rQ_{k+1} + \rS_{k+1} \le \half C_d^{-2N} L_{k+1}^{-\kappa(1+\theta)^{k+1}}.
$$
Therefore,
$$
\bal
\rP_{k+1} &\le \half  C^{2N}_d L_{k+1}^{ 2Nd } \rP_k^{2}
+ \frac{1}{2} C_d^{-2N} L_{k+1}^{-\kappa(1+\theta)^{k+1}}
%
\le \frac{C_d^2N}{C_d^{4N}} L_{k+1}^{-\frac{2\kappa(1+\theta)^k}{\alpha} + 2Nd}
\\
&\le  C_d^{-2N} L_{k+1}^{-\kappa(1+\theta)^{k+1}},
\eal
$$
provided that $\frac{2\kappa}{\alpha} + 2Nd \ge \kappa$
(i.e., $\kappa\ge \frac{2Nd\alpha}{2 - \alpha}$)
and $\theta =\left(\frac{2}{\alpha} - \frac{2d}{\kappa}\right) - 1 >0$.
\qedhere

\vskip4mm

\emph{This marks the end of the fixed-energy multi-particle multi-scale analysis.}

\section{From fixed to variable energy: First approach}
\label{sec:SW.ETV}

Now we establish a fairly general relation between fixed-energy
probabilistic estimates on the Green functions and variable-energy bounds
for two disjoint finite volumes. It does not matter how the probabilistic input is obtained; in particular, the results of this section can be combined both with the MSA,
performed for each fixed
energy $E$ in a given interval $I\subset \DR$, and with the FMM (which always starts
as a fixed-energy analysis).

\subsection{Derivation of the first variable-energy bound}
\label{ssec:SW.ETV.FEMSA.to.VEMSA}

The results of this
subsection\footnote{I thank Tom Spencer and Sasha Sodin for stimulating discussions
of the works \cite{ETV10,ESS12}, and Ivan Veseli\'c for a fruitful
exchange on this subject.}
are based on a straightforward adaptation
of the techniques developed by Elgart \emph{et al.} \cite{ETV10}.
In fact, the assertion of Theorem \ref{thm:SW.ETV} below is a mere
encapsulation of an argument from \cite{ETV10} in a statement involving
four parameters which can be adapted in various ways to particular models.

It is convenient to assume that $|I|=1$, so the interval $I$ with the Lebesgue measure
$\mes(\cdot)$ is a probability space, and so is the product space $(\Om\times I, \DP\times \mes)$.
Given $L\in\DN$ and points $\Bx,\By\in\bcZN$, set for brevity
\be\label{eq:def.Mxy}
\BF_{\Bx,\By}(E)= |\BG_{\bball_L}(\Bx,\By;E)|,
\;
\BF_{\Bx}(E)=\max_{\By\in\pt^- \bball_L(\Bx)} \BF_{\Bx,\By}(E),
\ee
and introduce the subsets of $I$ parameterized by $a>0$:
\be
\csE_{\Bx,\By}(a) = \{E\in I:\, \BF_{\Bx,\By}(E) \ge a \},\;\;
\csE_{\Bx}(a) = \{E\in I:\, \BF_{\Bx}(E) \ge a \}.
\ee
(The $L$-dependence will be often omitted for brevity.)


\begin{theorem}\label{thm:SW.ETV}
Let $L\ge 0$, $\Bx\in\bcZN$, $\By\in\pt^-\bball_L(\Bx)$. Let $\{\lam_j\}_{j=1}^{N}$  be the eigenvalues
of the operator $\BH_{\bball_L(\Bx)}(\om)$ and $I\subset\DR$ an interval of unit length.
Let be given numbers $a,b,c, \cP_L>0$
such that
\be\label{eq:cond.a.b.c}
b \le \min\left\{ |\bball_L(\Bx)|^{-1} ac^{2}, \, c \right\},
\ee
and for all $E\in I$
\be\label{eq:thm.fixed.E.intervals}
\pr{ \csE_{\Bx}(a)} \equiv \pr{ \BF_{\Bx}(E) \ge a } \le \cP_L.
\ee
There is an event
$\cB_{\Bx}(b)$ with $\pr{\cB_{\Bx}(b)}\le b^{-1} \cP_L$ such that
$\forall$ $\om\not\in\cB_{\Bx}(b)$, the  set
$$
\csE_{\Bx}(2a) = \csE_{\Bx}(2a;\om) = \Big\{ E: \, \BF_{\Bx}(E) \ge 2 a \Big\}
$$
is contained in a union of intervals
$\cup_{j=1}^N I_j$,
$I_j := \{E: |E-\lam_j| \le c \}$, $\lam_j\in I$.
\end{theorem}
\proof
Consider the following events parameterized by $b>0$:
\be\label{def:cS.cE.cB}
\bal
\cB_{\Bx}(b)
&= \myset{ \om\in \Om:\, \mes( \csE_{\Bx}(a) ) > b }.
\eal
\ee
Apply Chebyshev's inequality and the Fubini theorem
combined with \eqref{eq:thm.fixed.E.intervals}:
\begin{align}
\pr{ \cB_{\Bx}(b) }
&\le b^{-1} \esm{ \mes( \csE_{\Bx}(a) )  }
\nonumber
\\
& = b^{-1} \int_I \, dE\,\, \esm{ \one_{\{\BF_{\Bx}(E) \ge a \}} }
\label{eq:prob.cB}
 \le b^{-1} \cP(L).
\end{align}
Now fix any $\om\not\in \cB_{\Bx}(b)$, so that
$
\mes( \csE_{\Bx}(a); \om ) \le b.
$
There is a subset $\{\lam_j \}_{j=1}^{N'}$ of the EVs of the operator
$\BH_{\bball_L(\Bx)}$ such that
the Green function $E\mapsto \BG_{\bball_L(\Bx)}(\Bx,\By;E)$  reads as a rational function
(below we remove the vanishing terms, if any)
\be\label{eq:GF.rational}
f: E \mapsto \BG_{\bball_L(\Bx)}(\Bx,\By;E) =: \sum_{j=1}^{N'} \frac{\kappa_j}{\lam_j - E},
\;\; N' \le N := |\bball_L(\Bx)|;
\ee
here $\kappa_j = \kappa_j(\Bx,\By)\ne 0$ and
$\sum_j |\kappa_j| \le \sum_i |\psi_i(\Bx)\psi_i(\By)| \le N$. Let
$$
\bal
\csR(2c) &= \big\{ \lam\in\DR:\;  \min_j \, |\lam_j - \lam| \ge 2 c \big\},
\\
\csR(c) &= \big\{ \lam\in\DR:\;  \min_j \, |\lam_j - \lam| \ge  c \big\},
\quad c>0.
\eal
$$
Observe that, with $0<b\le c$,  $\csA_b:=\{E:\, \dist(E, \csR(2c)) < b \}\subset\csR(c)$,
hence, the set $\csA_b$ is a union of open sub-intervals at distance $\ge c$ from the spectrum,
and on each
sub-interval one has $|f'(E)| \le N c^{-2}.
$
Let us show
by contraposition
that, with $\om\not\in \cB_{\Bx,\By}(b)$,
$$
\myset{E:\, |\BG_{\bball_L}(\Bx,\By;E)| \ge 2 a } \cap \csR(c)=\varnothing.
$$
Assume otherwise, pick any point $\lam^*$ in the non-empty set in the LHS,
and let $J := \{ E': \, |E' -\lam^*|\le b\}\subset\csA_b\subset\csR(c)$.
Then for any
$E\in J$
one has, by \eqref{eq:cond.a.b.c},
$$
\bal
|f(E)| & \ge |f(\lam^*)| - | J| \sup_{E'\in J} |f'(E')|
%
> 2 a - N c^{-2} \cdot b  \ge  a,
\eal
$$
so
$J \subset \csE_{\Bx,\By}(a)$  and
$\mes(\csE_{\Bx,\By}(a) ) \ge \mes(J) = 2 b > b$, contrary to the choice of $\om$.
Since the set $\csR(c)$ is independent of $\By$, the assertion follows from \eqref{eq:prob.cB}.
\qedhere

\smallskip

Taking into account the FEMSA bound obtained in Section \ref{sec:FEMSA},
$\cP_{L_k} = \rP_k \le L_k^{-\kappa(1+\theta)^k}$,
one can set, e.g., with $\alpha=\frac{3}{2}$ and $\kappa > 6Nd$,
for $L \in\{L_k, k\ge 0\}$:
\be\label{eq:choice.a.b.c.}
a(L_k) = L_k^{-\frac{3\kappa}{5}(1+\theta)^k}, \;\;
b(L_k) = L_k^{-\frac{\kappa}{5}(1+\theta)^k}, \;\;
c(L_k) = L_k^{-\left( \frac{\kappa}{5} - \frac{d}{2} \right) (1+\theta)^k}.
\ee
Since $\kappa>6Nd$, one has $c(L_k) < L_k^{ - \frac{\kappa}{30}(1+\theta)^k}$.

\begin{theorem}\label{thm:2vol.VEMSA.ETV}
If the property {\rm \Wone} holds true, then for any pair of
separable balls $\bball_{L_k}(\Bx)$, $\bball_{L_k}(\By)$, $k\ge 0$, and a bounded
interval $I\subset\DR$,
one has
$$
\pr{\exists\, E\in I:\;
\min\{ \BF_\Bx(E), \BF_\By(E)\} \ge L_k^{-\frac{3\kappa}{5}(1+\theta)^k} }\le
C\, L_k^{- \frac{\delta\kappa}{30}(1+\theta)^k + (2N+1)d}.
$$
\end{theorem}

\proof
Introduce the event $\cB_b(\Bx)$, $b=b(L_k)$,
relative to the ball $\bball_{L_k}(\Bx)$ and defined as in Theorem
\ref{thm:SW.ETV}; similarly, define the event $\cB_b(\By)$ relative to the ball $\bball_{L_k}(\By)$
and set $\cB_b = \cB_b(\Bx)\cup \cB_b(\By)$. Further, let
$\csE_{\Bx} = \{E\in I:\,\BF_\Bx \ge 2a(L_k) \}$,
$\csE_{\By} = \{E\in I:\,\BF_\By \ge 2a(L_k) \}$, as in Theorem \ref{thm:SW.ETV}.
For any $\om\not\in\cB_b$,
$$
\dist\Big(\csE_{\Bx}, \Sigma(\BH_{\bball_{L_k}(\Bx)}) \Big) \le 2c(L_k), \quad
\dist\Big(\csE_{\By}, \Sigma(\BH_{\bball_{L_k}(\By)}) \Big) \le 2c(L_k).
$$
Therefore, applying Theorem \ref{thm:W2}, we obtain:
\be
\bal
\pr{ \csE_{\Bx} \cap \csE_{\By} \ne \varnothing } & =
\pr{ \dist\left( \csE_{\Bx}, \csE_{\By} \right) = 0 }
\\
&\le \pr{\cB_b} +
\pr{
\dist\Big(\Sigma(\BH_{\bball_{L_k}(\Bx)}), \Sigma(\BH_{\bball_{L_k}(\By)}) \Big) \le 4c(L_k) }
\\
& \le {L}_k^{- \frac{4\kappa}{5}(1+\theta)^k}
  + \Const\, {L}_k^{(2N+1)d} (c(L_k))^\delta
\\
& \le  \Const\, {L}_k^{- \frac{\delta\kappa}{30}(1+\theta)^k + (2N+1)d}.
\qedhere
\eal
\ee

\subsection{Spectral localization}

The assertion of Theorem \ref{thm:2vol.VEMSA.ETV} has a structure
similar to that of the MSA bound from the work by von Dreifus and Klein \cite{DK89}.
More precisely, it guarantees a decay rate of Green functions slower than exponential,
but faster than any power-law. 
The main difference is that one can rule out, with high  probability,
only the pairs of singular balls which are separable (and not just disjoint).
It is not difficult to adapt the well-known argument from
\cite{DK89} and prove that with probability one,
all polynomially bounded solutions to the equation $H(\om) \psi = E\psi$ are in fact
square-summable. The latter property requires a Shnol--Simon type result
on spectrally a.e. polynomial boundedness of generalized eigenfunctions.
It will follow independently by RAGE (Ruelle--Amrein--Georgescu--Enss) theorems
(see a detailed discussion along with a bibliography, e.g., in \cite{CyFKS87}) from
the dynamical localization proven in Section \ref{sec:MSA.to.DL}.

\vskip4mm
\section{From fixed to variable energy: Second approach}
\label{sec:SW.PCT}

\subsection{The spectral reduction}

The next result is a notational adaptation of Theorem 7.1 from \cite{C12b},
formulated there for single-particle Hamiltonians.

\begin{theorem}\label{thm:SW.CPT}
Let be given a ball $\bball_L(\Bx)$, $L\ge 1$, and
numbers $a(L)$, $b(L)$, $c(L)$, $\cP_L>0$ obeying \eqref{eq:cond.a.b.c} and
such that, for some interval $I$, all $E\in I$,
\be\label{eq:thm.fixed.E.intervals.3}
\pr{ \BF_{\Bx}(E) \ge a } \le \cP_L.
\ee
Set $K=|\bball_L(\Bx)|$.
Then the following properties hold true:
\begin{enumerate}[\rm(A)]
\item
For any $b\ge \cP_L$ there exists an event $\cB_b$
such that $\pr{\cB_b}\le b^{-1}\cP_L$
and for any $\om\not\in\cB_p$ the set of energies
$$
\csE_{\Bx}(a) = \csE_{\Bx}(a;\om) = \{ \BF_{\Bx}(E) \ge a\} \cap I
$$
is covered by $\tK< 3K^2$ intervals $J_i = [E^-_i,E^+_i]$,
of total length
$\sum_i |J_i|\le b$.
  \item The endpoints $E^\pm_i$ are determined by the functions
  $E\mapsto \langle \one_\Bx| (\BH_{\bball_L(\Bu)} - E)^{-1} | \one_\By \rangle$
  in such a way that, for the one-parameter
  family $\BA(t) := \BH_{\bball_L(\Bu)} + t\one$, the endpoints $E^\pm_i(t)$ for
  the operators $\BA(t)$ (replacing $\BH_{\bball_L(\Bu)}$) have the form
$$
   E^\pm_i (t) = E^\pm_i (0) + t, \quad t\in\DR.
$$

\end{enumerate}

\end{theorem}

\proof
(A) Fix a point $\By\in\pt^-\bball_L(\Bx)$ and consider the rational function
$$
f_\By: E \mapsto \sum_{i=1}^K \frac{\kappa_i}{ \lam_i - E}
:= \sum_{i=1}^K \frac{\Bpsi_i(x) \, \Bpsi_i(\By)}{ \lam_i - E}.
$$
Its derivative has the form
$$
f'_\By(E)  = \sum_{i=1}^K \frac{-\kappa_i}{ (\lam_i - E)^2}
=: \frac{\csP(E)}{ \csQ(E)}, \;\; \deg \, \csP \le 2K-2,
$$
and has $\le 2K-2$ zeros and $\le K$ poles, so $f_\By$ has $\le 3K-1$ intervals of monotonicity
$I_{i,\By}$, and the total number of monotonicity intervals of all functions
$\{f_\By, \By\in\pt^-\bball_L(\Bx)\}$ is bounded by
$\tK\le|\pt^-\bball_L(\Bx)|(3K-1)\le|\bball_L(\Bx)|(3K-1)<3K^2$, so
$$
\cup_{\By\in\pt^-\ball_L(\Bx)}
\{E:\; f_\By(E) \ge a\} = \cup_{i=1}^{K} J_i, \quad
J_i = [E^-_i,E^+_i] \subset I,
$$
where, obviously, $\sum_i |J_i|\le\mes\, \{E:\; \BF_\Bx(E) \ge a\}$.
\par\vskip2mm
\noindent
(B)  Consider a one-parameter operator family
$
\BA(t) = \BH_{\bball_L(\Bu)}(\om)  + t\one.
$
All these operators share common
eigenvectors; the latter determine the coefficients $\kappa_i$, so one can choose
eigenfunctions $\psi_i(t)$ constant in $t$ and obtain $\kappa_i(t) = \kappa_i(0)$.
The eigenvalues of operators $\BA(t)$ have the form $\lam_i(t) = \lam_i(0)+t$. We conclude
that the Green functions, with fixed $\Bx$ and $\By$, have the form
$f_{\Bx,\By}(E;t) = f_{\Bx,\By}(E-t;0)$, so that the intervals $J_i(t)$ have indeed the form
$J_i(t) = [E^-_i + t, E^+_i + t]$.
\qedhere

\begin{theorem}\label{thm:SW.2vol.PCT}
Assume the property {\rm \Wtwo}.  Let be given numbers $a>0$ and $\cP_L\in[0,1]$ such that  \eqref{eq:thm.fixed.E.intervals.3} holds true for all $\Bx\in\bcZN$, then for any pair of
weakly separable (e.g., distant) balls $\bball_L(\Bx)$, $\bball_L(\By)$ and for any $b \ge \cP_L$,
one has
\be\label{eq:thm.SW.2vol.PCT}
\pr{\exists\, E\in I:\; \min\{ \BF_\Bx(E), \BF_\By(E)\} \ge a}
\le 2b^{-1} \cP_L +  \htil_L(4b),
\ee
where
$$
\htil_L(s) := K^2 C L^{A} s^{B} + C' L^{A'} s^{B'}, \quad
K=\max\{|\bball_L(\Bx)|, |\bball_L(\By)|\}.
$$
\end{theorem}

\proof
Fix $b>0$ (clearly, $b< \cP_L$ gives rise to a trivial bound $\pr{}< 1$).
Define the event $\cB_b(\Bx)$ relative to the ball $\bball_L(\Bx)$ as in Theorem
\ref{thm:SW.CPT}. Similarly, define the event $\cB_b(\By)$ relative to the ball $\bball_L(\By)$,
and let $\cB_b = \cB_b(\Bx)\cup\cB_b(\By)$. Further, let $\cS_{\Bx,\By}$ be the event
figuring in the LHS of \eqref{eq:thm.SW.2vol.PCT} and note that
$$
\pr{ \cS_{\Bx,\By}} \le \pr{\cB_b} + \pr{ \cS_{\Bx,\By} \cap \cB_b^\rc}
\le 2b^{-1} \cP_L + \pr{ \cS_{\Bx,\By} \cap \cB_b^\rc}.
$$
It remains to assess $\pr{ \cS_{\Bx,\By} \cap \cB_b^\rc}$.
By Definition \ref{def:ws}, there is a ball $\Lam\subset\cZ$
of diameter $R\le 2NL$ and index subsets $\cJ_1, \cJ_2\subset[1,N]$ such that
$n_1:=|\cJ_1|>|\cJ_2| =: n_2$ and
\begin{equation}\label{eq:cond.WS.again}
\left\{\begin{array}{l}
 \PPi_{\cJ_1} \bball^{(N)}_L(\Bx) \cup \PPi_{\cJ_2} \bball^{(N)}_L(\By) \;  \subseteq \Lam,
\\
\PPi_{\cJ^c_1} \bball^{(N)}_L(\Bx) \cap \Lam = \varnothing,
\\
\PPi_{\cJ^c_2} \bball^{(N)}_L(\By) \cap \Lam = \varnothing.
\end{array}
\right.
\end{equation}
Consider the random variables
$$
\xi_{\Lam}(\om) := |\Lam|^{-1}\sum_{z\in\Lam} V(z;\om),\quad
\eta_z(\om) := V(z;\om) - \xi_\Bx(\om), \; z\in\Lam.
$$
Denote the $\sigma$-algebra generated by the random variables
$\{\eta_z, z\in\Lam; V(u,\cdot), u\not\in\Lam\}$ by $\fF_{\eta,\Lam}$.
Consider the conditional probability distribution function
$$
F_{\xi_{\Lam}}(t\,|\, \fF_{\eta,\Lam})
= \pr{ \xi_\Lam \le t \,|\, \fF_{\eta,\Lam}}
$$
and its continuity modulus
$$
\nu_{\xi_{\Lam}}( s\,|\, \fF_{\eta,\Lam}) =
\sup_{t\in\DR}\; \essup  \;
\left( F_{\xi_{\Lam}}(t+s\,|\, \fF_{\eta,\Lam})
   - F_{\xi_{\Lam}}(t\,|\, \fF_{\eta,\Lam})\right).
$$
Owing to the assumption \Wtwo, for some $C,C', A,A',B,B'\in(0,+\infty)$
we have
\be\label{eq:cond.PCT}
\forall\, s\in[0,1] \qquad
\pr{ \nu_{\xi_{\Lam}}( s\,|\, \fF_{\eta,\Lam}) > C L^{A} s^B }
\le C' L^{A'} s^{B'}.
\ee
Using the representation $V(z;\om) = \xi_{\Lam}(\om)\one + \eta_z(\om)$
in the set $\Lam$, introduce the respective operator decomposition
\be\label{eq:H.A.xi}
\BH_{\ball_L(\Bx)}(\om) =  n_1\xi_{\Lam}(\om)\one + \BA_\Bx(\om),
\ee
where the operator $\BA_\Bx(\om)$ is non-random, conditional on $\fF_{\eta,\Lam}$.
For any $\om\not\in\cB_b(\Bx)$, the energies $E$ where
$\BF_{\Bx}(E)\ge a$
are covered by intervals $J_{i,\Bx}$,
with $\sum_i |J_{i,\Bx}|\le b$.
Combining \eqref{eq:H.A.xi} and the assertion (B) of Theorem \ref{thm:SW.CPT}, we can write
$$
J_{i,\Bx}(\om) =
[\tE^-_{i,\Bx}(\om) + n_1\xi_{\Lam}(\om), \; \tE^+_{i,\Bx}(\om)
   + n_1 \xi_{\Lam}x(\om)]
$$
where $\tE^{\pm}_{i,\Bx}(\om)$ are $\fF_{\eta,\Lam}$-measurable.
For any $\om\not\in\cB_b(\By)$,
the energies $E$ where $\BF_\By(E)\ge a$ are covered by intervals
$J_{i,\By}$, also obeying
$\sum_i |J_{i,\By}|\le b$. As above,
$$
J_{i,\By}(\om) =
[\tE^-_{i,\By}(\om) + n_2\xi_{\Lam}(\om), \; \tE^+_{i,\By}(\om)
   + n_2 \xi_{\Lam}(\om)]
$$
where $\tE^{\pm}_{i,\By}(\om)$ are also $\fF_{\eta,\Lam}$-measurable.

Let $\cB_b = \cB_b(\Bx)\cup\cB_b(\By)$ and set
$\eps_{i,\Bx} = |J_{i,\Bx}|$, $\eps_{j,\By} = |J_{j,\By}|$.
For $\om\in\cB_b^\rc$, one has $0\le \eps_{i,\Bx}, \eps_{i,\By}\le 2b$, so
$$
\bal
\{\om:\, J_{i,\Bx}\cap J_{j,\By} \ne \varnothing\} \cap \cB_b^\rc
&\subset
\left\{ \big| \tE^-_{i,\Bx}(\om) - \tE^-_{j,\By}(\om) \big| \le \eps_{i,\Bx} + \eps_{j,\By}
\right\} \cap \cB_b^\rc
\\
& \subset \left\{ \big| (n_1-n_2)\xi_{\Lam}(\om) - \tE_{i,j}(\om) \big| \le 2b + 2b
\right\}
\eal
$$
where $\tE_{i,j}(\om)$ is $\fF_{\eta,\Lam}$-measurable.
Set $\tn = n_1-n_2 (\ge 1)$.
By  \eqref{eq:cond.PCT},
$$
\bal
 \pr{ \big| \tn \xi_{\Lam}(\om) - \tE_{i,j}(\om) \big| \le 4b }
&= \esm{ \pr{ \big| \tn\xi_{\Lam}(\om) - \tE_{i,j}(\om) \big| \le 4b \,|\, \fF_{\eta,\Lam} } }
\\
& \le \pr{ \nu_{\xi_{\Lam}}( 4b\,|\, \fF_{\eta,\Lam}) > C L^{A} (4b)^B }
+ C L^{A} (4b)^{B}.
\eal
$$
Observe that the first probability in the RHS does not depend upon
$i$ and $j$, so it suffices to count it only once.
Taking the sum over $(i, j)$,  we obtain the required bound:
$$
\bal
\pr{ \cS_{\Bx,\By} \cap \cB^\rc} &\le 2b^{-1} \cP_L +
 \pr{\om:\, \cup_{i,j} (J_{i,\Bx}\cap J_{j,\By}) \ne \varnothing}
\\
&  \le 2b^{-1} \cP_L + \htil_L(4b).
\qedhere
\eal
$$
\par
\vskip4mm

In particular, taking into account Theorem \ref{thm:MSA.main.fixed}, we can set, for $L=L_k$,
$$
a=a(L_k)=\eu^{-\gamma(m,L_k)L_k}, \;
\cP_{L_k} = L_k^{-\kappa(1+\theta)^k}, \;
b=b(L_k) =L_k^{-\frac{\kappa}{2}(1+\theta)^k}.
$$
These settings give rise to the following corollary of Theorem \ref{thm:SW.2vol.PCT}:

\begin{theorem}\label{thm:2vol.VEMSA.PCT}
Assume the property {\rm\Wtwo}.
If there is an integer $L_0\ge 1$ and numbers $m\ge 1$, $\alpha\in(1,2)$ such that
$$
\textstyle
\min \{\cP_0, \rQ_0\} \le C_d^{-2N} L_{0}^{-\kappa}, \; \kappa > \frac{2\alpha Nd}{2 - \alpha},
$$
then for some $C,a,c>0$ and all $k\ge 0$, any interval $I\subset\DR$ with $|I|\le 1$
and any pair of distant balls $\bball_{L_k}(\Bx)$, $\bball_{L_k}(\By)$, the
following bound holds true:
\be\label{eq:thm.VEMPMSA.ind}
\pr{E\in I:\; \bball_{L_k}(\Bx) \text{ and $\bball_{L_k}(\By)$ are {\rm\EmS} } }
\le C\, \eu^{-a \ln^{1+c} L_k}.
\ee
\end{theorem}

\subsection{On the validity of the assumption \eqref{eq:cond.PCT}}
\label{ssec:validity.W2}

First of all, recall that, by an elementary result on Gaussian distributions, if  $V:\cZ\times\Om\to\DR$ is an IID Gaussian field, say, with zero mean and unit variance,
the sample average $\xi_x$ of the sample $\{V(z;\om), z\in\ball_L(x)\}$ is independent of
the sigma-algebra generated by the "fluctuations" $\eta_z(\om)$; moreover, it has Gaussian distribution $\cN(0, |\ball_L(x)|)$ and admits a probability density
with $\|p_{\xi_x}\|_\infty\le \frac{1}{\sqrt{2\pi}}\,|\ball_L(x)|^{1/2}$. In this particular case,
Eqn \eqref{eq:cond.PCT} can be replaced by a stronger, deterministic bound: the conditional
continuity modulus $\nu_{\xi_x}( s\,|\, \fF_x)$ is actually independent of the condition and is
a.s. bounded by $\|p_{\xi_x}\|_\infty \cdot s$.

Such a situation is rather exceptional, as shows the example of two IID random variables
$V_1(\om), V_2(\om)$ with uniform distribution ${\rm Unif}([0,1])$. Indeed, in this case
$\xi := (V_1+V_2)/2$, $\eta = (V_1-V_2)/2$ and the distribution of
$\xi$ conditional on $\eta$ is uniform on the interval $I_\eta$ of length $O(1-|\eta|)$, hence,
with constant density $O(\big|1 - |\eta|\big|^{-1})$, for $|\eta|<1$; for $\eta = \pm 1$,
this distribution is concentrated on a single point. However, this example shows also
how such a difficulty can be bypassed: excessively "singular" conditional distrubutions
of the sample mean $\xi$ occur only for a set of conditions having a small probability.
Using this simple idea, Gaume \cite{G10}, in the framework of his PhD project,
established the property \eqref{eq:cond.PCT} for
IID random fields with piecewise constant marginal \emph{probability density}. By standard
approximation arguments, it can be easily extended to piecewise Lipshitz (or H\"{o}lder)
continuous \emph{densities}, which is sufficient for
most physically relevant applications.
We believe that some variant of the property \eqref{eq:cond.PCT},
perhaps weaker but still sufficient for the purposes of the MSA, holds true
in a larger class of IID random fields.

\subsection{Exponential spectral localization}

The assertion of Theorem \ref{thm:2vol.VEMSA.PCT} has the same form as in the
conventional MSA bound going back to the work by von Dreifus and Klein \cite{DK89}
(actually, even slightly stronger); therefore, the same argument as in
\cite{DK89} (having its roots in \cite{FMSS85}) applies and proves that with probability one,
all polynomially bounded solutions to the equation $H(\om) \psi = E\psi$ are in fact
decaying exponentially fast at infinity, thus the operator $H(\om)$
has a.s. pure point spectrum. The latter property requires a Shnol--Simon type result
on spectrally a.e. polynomial boundedness of generalized eigenfunctions;
it will also follow by RAGE theorems from
the dynamical localization proven in Section \ref{sec:MSA.to.DL}.

\section{From MSA to the strong dynamical localization}
\label{sec:MSA.to.DL}

This section is a straightforward adaptation of Section 8 from \cite{C12b}. Formulated
in the most general form, the results of this section apply indifferently
to single- and multi-particle Hamiltonians, and the particularity of the latter
resides in the geometry of pairs of finite balls involved.

Recall that the first rigorous derivations of the dynamical localization from
MSA-type probabilistic bounds on
the Green functions have been obtained by Germinet--De Bi\`{e}vre \cite{GD98} and
Damanik--Stollmann \cite{DS01}.
Germinet and Klein \cite{GK01} proposed a shorter proof, which
can be further simplified in the context of finite-volume
operators (which merely should have compact resolvent),
and essentially reduced to an elementary application of the Bessel inequality.

In applications to $N$-particle systems on a graph obeying
the growth condition \eqref{eq:ball.growth},
the parameter $D$ below has to be replaced by $Nd$.

\subsection{EF correlators in finite balls}
\label{sec:MSA.to.DL.finite}

Given an interval $I\subset\DR$, denote by $\csB_1(I)$ the set of all Borel functions $\phi:\DR\to\DC$ with
$\supp\,\phi\subset I$ and $\|\phi\|_\infty \le 1$.

\begin{theorem}\label{thm:GK}
 Fix an integer $L\in\DN^*$ and assume that  the following bound holds for any pair of disjoint balls $\ball_L(x), \ball_L(y)$ and some quantity $\varsigma(L)>0$:
$$
\pr{ \exists\, E\in I:\, \text{ $\ball_L(x)$ and $\ball_L(y)$ are $(E,m)${\rm-S}} } \le \varsigma(L).
$$
Then for any $x,y\in\cZ$ with $\rd(x,y)> 2L+1$, any finite connected subgraph (of $\cZ$)
$\cG\supset\ball_L(x) \cup \ball_L(y)$ and any Borel function $\phi\in\csB_1(I)$
\be\label{eq:thm.MSA.to.DL}
\esm{ \big|\langle\one_x | \phi(H_\cG(\om)) | \one_y \rangle \big| }
\le 4 \eu^{-mL} + \varsigma(L).
\ee
\end{theorem}

\proof Fix points $x,y\in\cZ$ with $\rd(x,y)> 2L+1$ and a graph
$\cG\supset\ball_L(x) \cup \ball_L(y)$. The operator $H_\cG(\om)$
has a finite orthonormal eigenbasis $\{\psi_i\}$ with respective eigenvalues
$\{\lam_i\}$. Let $\BbS = \pt \ball_L(x) \cup \pt \ball_L(y)$
(recall: this is a set of \emph{pairs}
$(u,u')$); note that $|\BbS|\le 2C_D^2L^D$, by \eqref{eq:ball.growth}. Suppose that for some $\om$, for each $i$ there is $z_i\in \{x,y\}$ such that
$\ball_L(z_i)$ is $(\lam_i,m)$-NS; let $\{v_i \}= \{x,y\}\setminus \{z_i\}$.
Denote
$\mu_{x,y}(\phi) = \big|\langle\one_x | \phi(H_\cG(\om)) | \one_y \rangle \big|$,
with $\mu_{x,y}(\phi)\le 1$.
Then by the GRI for the eigenfunctions,
%
\begin{align*}
 \mu_{x,y}(\phi)
& \le \|\phi\|_\infty \, \sum_{\lam_i \in I} |\psi_i(x) \psi_i(y)|
\le \sum_{\lam_i \in I} |\psi_i(z_i) \psi_i(v_i)|
\\
& \le \sum_{\lam_i \in I} |\psi_i(v_i)| \, \eu^{-mL} (C_D^2L^D)^{-1}
      \sum_{(u,u')\in\pt \ball_L(z_i)} |\psi_i(u)|
\qquad\qquad\qquad\qquad\qquad\qquad
\\
& \le \eu^{-mL} \sum_{\lam_i \in I} \; (C_D^2L^D)^{-1} \sum_{(u,u')\in \BbS}
      |\psi_i(u)| \left(|\psi_i(x)| + |\psi_i(y)|  \right)
\\
&\le \eu^{-mL} \frac{|\BbS|}{ C_D^2L^D} \, \max_{u\in\cG} \sum_{\lam_i \in I}
    \half \left( |\psi_i(u)|^2 + |\psi_i(x)|^2 + |\psi_i(y)|^2 \right)
\\
\intertext{(using Bessel's inequality and $|\BbS|\le 2C_D^2L^D$)}
& \le  \eu^{-mL} \,2 \, \max_{u\in\cG}
       \left( 2\|\one_u\|^2 + \|\one_x\|^2 + \|\one_y\|^2 \right)
%
= 4 \eu^{-mL}.
\end{align*}
Denote
$\cS_L = \myset{\exists\, E\in I:\, \text{ $\ball_L(x)$ and $\ball_L(y)$ are $(E,m)$-S}}$,
with $\pr{\cS_L}\le \varsigma(L)$, by assumption. Now we conclude:
$$
\esm{ \mu_{x,y}(\phi) } = \esm{ \one_{\cS_L}\mu_{x,y}(\phi) }
   + \esm{ \one_{\cS^{\rc}_L}\mu_{x,y}(\phi) }  \le \varsigma(L) + 4\eu^{-mL}.
\qedhere
$$

\subsection{Dynamical localization on the entire graph}
\label{sec:MSA.to.DL.infinite}

For the reader's convenience, we repeat below a simple argument,
described in \cite{C12b} and employed earlier by Aizenman et al. \cite{A94,ASFH01}. The quantities
$\mu_{x,y}^{(H)}(\phi) = \langle \one_x \,|\, \phi(H)\,|\, \one_y\rangle$ defined, for example, for bounded continuous or Borel functions $\phi$, generate signed (i.e., not necessarily positive)
spectral measures associated with a self-adjoint operator $H$:
$$
\int d\mu_{x,y}^{(H)}(E) \, \phi(E) := \langle \one_x | \phi(H)| \one_y\rangle.
$$
In particular, we can consider, with $x,y,u\in\cZ$ fixed, measures
$\mu^k_{x,y}$ related to operators $H_{\ball_{L_k}(u)}$, for all $k\ge 0$,
as well as their counterparts $\mu_{x,y}$ for the operator $H$
on the entire graph $\cZ$. A sufficient condition for the vague convergence
$\mu^k_{x,y}\rightarrow \mu_{x,y}$ as $k\to\infty$ is the strong resolvent convergence
$H_{\ball_{L_k}(u)} \rightarrow H$. Such convergence is well-known to occur for a very large class
of operators, including (unbounded) Schr\"{o}dinger operators in Euclidean spaces
and their analogs on the so-called quantum graphs. Indeed, for (not necessarily bounded)
operators $H_n$ with a common core $\cD$ to converge to an operator $H$ with the same core,
it suffices that $H_k \psi \to H\psi$ strongly for any element $\psi\in\cD$ (cf. \cite{Kato}).
For finite-volume operators, one can usually find an appropriate core $\cD$ formed by compactly supported functions $\psi$; for finite-difference Hamiltonians on graphs (even unbounded,
e.g., for DSO with unbounded potentials) one can choose
as $\cD$ the subset of all functions with finite supports. On such functions,
$H_{\ball_{L_k}(u)}\psi \to H\psi$ as $k\to\infty$ (by stabilization),
therefore, the spectral measures converge vaguely:
$\mu^k_{x,y}\rightarrow \mu_{x,y}$. By Fatou lemma,
for any bounded Borel set $A\subset\DR$,
one has
$$
\left| \mu_{x,y}(A) \right| \le \liminf_{k\to\infty} \left| \mu^k_{x,y}(A) \right|
$$
(here $\left| \mu(A) \right| := \sup \{ \mu(\phi), \, \|\phi\|\le 1, \, \supp\, \phi\subset A\}$).
Taking the expectation and using the uniform upper bounds on EF correlators in finite balls,
we conclude that
\be
\esm{ \sup_{\phi\in\csB_1} \, \big| \langle \one_x | \phi(H(\om)) | \one_y \rangle \big| }
\le C \eu^{ - a \ln^{1+c} \rd(x,y)}
\ee
(using the inequality
$L_k^{-\kappa(1+\theta)^k} \le C \eu^{ - a \ln^{1+c} L_k}$, for some $C,a,c>0$).
In particular, with functions $\phi_t:\lam\mapsto \eu^{-it\lam}$, we obtain the strong dynamical localization property for the ensemble of random Hamiltonians $H(\om)$.

Taking into account Theorem \ref{eq:thm.VEMPMSA.ind}, we come to the following
sufficient conditions of strong $N$-particle dynamical localization in an interval
$I\subseteq\DR$:

\begin{theorem}\label{thm:from.VEMPMSA.to.DL.W1}
Assume {\rm\Wone} and suppose that, for some $N\ge 2$ and all $k\ge 0$,
\SSI{N,k} holds true. Then there are constants
$a, c\in(0,+\infty)$ such that for all $\Bx\in\bcZN$, some $C(\Bx)\in(0,+\infty)$ and all
$\By\in\bcZN$
\be
\esm{ \sup_{t\in\DR} \,
\big| \langle \one_\Bx | \eu^{-\ii t \BHN(\om)} P_I(\BHN(\om)) | \one_\By \rangle \big| }
\le C(\Bx) \,\eu^{ - a \ln^{1+c} \brho(\Bx,\By)}.
\ee
\end{theorem}

Here and below, $P_I(\BHN(\om))$ stands for the spectral projection of operator
$\BHN(\om)$ on the interval $I$. Recall that, formally, we assumed in Sections
\ref{sec:SW.ETV}--\ref{sec:MSA.to.DL}
the interval $I$ to be finite. In the case of a bounded random potential, the spectrum
of the operator is covered by a finite, non-random interval $I$, so that
$P_I(\BHN(\om)) = \BHN(\om)$.

\begin{theorem}\label{thm:from.VEMPMSA.to.DL.W2}
Assume {\rm\Wtwo} and suppose that, for some $N\ge 2$ and all $k\ge 0$,
\SSI{N,k} holds true. Then there are constants
$C, a, c\in(0,+\infty)$ such that for all $\Bx,\By\in\bcZN$
\be
\esm{ \sup_{t\in\DR} \,
\big| \langle \one_\Bx | \eu^{-\ii t \BHN(\om)} P_I(\BHN(\om)) | \one_\By \rangle \big| }
\le C \eu^{ - a \ln^{1+c} \brhoS(\Bx,\By)}.
\ee
\end{theorem}

\section{Appendix}

\subsection{Proof of Lemma \ref{lem:PITRONS}}
\label{ssec:proof.PITRONS}

The ball $\bball:=\bball_{L_k}(\Bu)$ is assumed PI, so there is a decomposition
of the configuration $\Bu$ into two non-interacting subconfigurations, $\Bu = (\Bu', \Bu'')$,
so that $\BU(\Bu) = \BU(\Bu') + \BU(\Bu'')$, and the operator
$\BH^{(N)}_{\bball}$ reads as follows:
\be\label{eq:Ham.algebraic.decomposable}
\BH_{\bball^{(N)}_{L_{k}}(\Bu)} =
\BH_{\bball^{(n')}_{L_{k}}(\Bu')} \otimes \one^{(n'')}
+ \one^{(n')} \otimes \,\BH_{\bball^{(n')}_{L_{k}}(\Bx'')}
\ee
thus its eigenvalues are the sums $E_{a,b}=\lam_a+\mu_b$, where
$\{\lam_a\}=\Sigma(\BH_{\bball_{L_{k}}(\Bu')})$ is the spectrum of
$\BH_{\bball_{L_{k}}(\Bu')}$ and, respectively,  $\{\mu_b\}=\Sigma(\BH_{\bball_{L_{k}}(\Bu'')})$.
Eigenvectors of $\BH^{(R)}_{\Bu,k}$ can be chosen in the form
$
\BPsi_{a,b} = \Bphi_a \otimes \Bpsi_b
$
where $\{\Bphi_a\}$ are eigenvectors of $\BH_{\bball^{(n')}_{L_{k}}(\Bu')}$ and
$\{\Bpsi_b\}$ are eigenvectors of $\BH_{\bball^{(n'')}_{L_{k}}(\Bu'')}$.
For each pair $(\lam_a,\mu_b)$, the non-resonance assumption $|E - (\lam_a + \mu_b)|\ge e^{L_k^\beta}$
reads as $|(E -\lam_a) - \mu_b)|\ge e^{L_k^\beta}$ and also as
$|(E -\mu_b) - \lam_a)|\ge e^{L_k^\beta}$.
Set
$\BG(E) = (\BH_{\bball_{L_{k}}^{(N)}(\Bu)} - E)^{-1}$; then we can write
\begin{eqnarray}
\BG(\Bu,\By; E) &=& \sum_{\lam_a} \sum_{\mu_b}
\frac{ \Bphi_a(\Bu') \Bphi_a(\By')\, \Bpsi_b(\Bu'') \Bpsi_b(\By'')\, }
{ (\lam_a + \mu_b) - E}
\\
\label{eq:a}
& =& \sum_{\lam_a} \BP'_a(\Bu',\By') \, \BG_{\bball_{L_k}(\Bu'')}(\Bu'',\By''; E-\lam_a)
\\
\label{eq:b}
& =& \sum_{\mu_b}  \BP''_b(\Bu'',\By'') \, \BG_{\bball_{L_k}(\Bu')}(\Bu',\By'; E-\mu_b),
\end{eqnarray}
By assumption, for all $\mu_b\in\sigma(\BH^{(R)}_{\bball_{L_k}(\Bu'')})$, the projection ball
$\bball_{L_k}(\Bu')$ is $(\mu_b,m)$-NS, and for all
$\lam_a\in\sigma(\BH^{(R)}_{\bball_{L_k}(\Bu')})$, the projection ball
$\bball_{L_k}(\Bu'')$ is $(\lam_a,m)$-NS.
%
%

For any $\By\in\pt^- \bball_{L_k}(\Bu)$, either
$\rho(\Bu',\By') = L_k$, in which case we infer from  \eqref{eq:b},
combined with $(\mu_b,m)$-NS property of the ball $\bball_{L_k}(\Bu')$, that
\be
\big| \BG(\Bu,\By; E) \big| \le |\bball_{L_k}(\Bu'')|\,
e^{-\gamma(m,L_k,N-1)L_k + 2L_k^{\beta}}
\ee
or $\rho(\Bu'',\By'') = L_k$, and then we have by
\eqref{eq:a}
$$
\big| \BG(\Bu,\By; E) \big| \le |\bball_{L_k}(\Bu')|\,
e^{-\gamma(m,L_k,N-1)L_k + 2L_k^{\beta}}.
$$
In either case, the LHS is bounded by
$$
\ba
\exp\left( -m(1+L_k^{-\tau})^{\hN-(N-1)+1} L_k + 2L_k^{\beta} + \Const \ln L_k \right)
< \half e^{ -\gamma(m,L_k,N)},
\ea
$$
for $L_0$ large enough, since $m\ge 1$ and
$$
L_k^{1-\tau} \equiv L_k^{7/8}
\gg L_k^{1/2} + \Const \ln L_k \equiv L_k^{\beta} + \Const \ln L_k \, .
$$
\qed

\section*{Acknowledgements}

I  thank Tom Spencer and the Institute for Advanced Study, Princeton,
for their hospitality
during my visit to the IAS in March, 2012, and for numerous stimulating discussions;
Sasha Sodin for fruitful discussions of the works \cite{ETV10}--\cite{ESS12}; G\"{u}nter Stolz,
Yulia Karpeshina, Roman Shterenberg and the University of Alabama at Birmingham for their
hospitality during my visit to the UAB in March, 2012, and for numerous  discussions; Ivan Veseli\'c for a fruitful discussion of the paper \cite{ETV10}.


\end{document}